\documentclass[journal]{IEEEtran}
\usepackage{amsthm,amsmath,amsfonts}
\usepackage{algorithmic}
\usepackage{array}
\usepackage[caption=false,font=normalsize,labelfont=sf,textfont=sf]{subfig}
\usepackage{textcomp}
\usepackage{stfloats}
\usepackage{url}
\usepackage{verbatim}
\usepackage{graphicx}
\usepackage{amssymb}
\usepackage{booktabs}
\usepackage{algorithm}  
\usepackage{cite}
\usepackage{epstopdf}
\usepackage{textcomp}
\usepackage{xcolor}
\usepackage{bm}
\usepackage{setspace}
\usepackage{balance}
\allowdisplaybreaks
\def\BibTeX{{\rm B\kern-.05em{\sc i\kern-.025em b}\kern-.08em
    T\kern-.1667em\lower.7ex\hbox{E}\kern-.125emX}}
\usepackage{balance}
\begin{document}
\title{Joint Discrete Antenna Positioning and Beamforming Optimization in Movable Antenna Enabled Full-Duplex ISAC Networks}
\author{Zhendong Li, Jianle Ba, Zhou Su, Haixia Peng, Yuntao Wang, Wen Chen, and Qingqing Wu

\thanks{Zhendong Li, Jianle Ba and Haixia Peng are with the School of Information and Communication Engineering, Xi'an Jiaotong University, Xi'an 710049, China (email: lizhendong@xjtu.edu.cn, 1650376377@stu.xjtu.edu.cn, haixia.peng@xjtu.edu.cn). }
\thanks{Zhou Su and Yuntao Wang are with the School of Cyber Science and Engineering, Xi'an Jiaotong University, Xi'an 710049, China (email: zhousu@ieee.org, yuntao.wang@xjtu.edu.cn).}
\thanks{Wen Chen and Qingqing Wu are with the Department of Electronic Engineering, Shanghai Jiao Tong University, Shanghai 200240, China (e-mail: wenchen@sjtu.edu.cn, qingqingwu@sjtu.edu.cn). }
\thanks{(\emph{Corresponding author: Zhou Su.})}}


\maketitle

\begin{abstract}
In this paper, we propose a full-duplex integrated sensing and communication (ISAC) system enabled by a movable antenna (MA). By leveraging the characteristic of MA that can increase the spatial diversity gain, the performance of the system can be enhanced. We formulate a problem of minimizing the total transmit power consumption via jointly optimizing the discrete position of MA elements, beamforming vectors, sensing signal covariance matrix and user transmit power. Given the significant coupling of optimization variables, the formulated problem presents a non-convex optimization challenge that poses difficulties for direct resolution. To address this challenging issue, the discrete binary particle swarm optimization (BPSO) algorithm framework is employed to solve the formulated problem. Specifically, the discrete positions of MA elements are first obtained by iteratively solving the fitness function. The difference-of-convex (DC) programming and successive convex approximation (SCA) are used to handle non-convex and rank-1 terms in the fitness function. Once the BPSO iteration is complete, the discrete positions of MA elements can be determined, and we can obtain the solutions for beamforming vectors, sensing signal covariance matrix and user transmit power. Numerical results demonstrate the superiority of the proposed system in reducing the total transmit power consumption compared with fixed antenna arrays.
\end{abstract}


\section{Introduction}
\definecolor{BLUE}{RGB}{0,0,255} 

\IEEEPARstart{{W}}{ith} the advancement of technology, the demand for efficient and reliable sensing and communication is increasingly high. For instance, in the Internet-of-vehicles (IoV), the capability to simultaneously offer vehicle positioning, speed monitoring and collision warning is essential to enhance road safety and traffic efficiency. To enhance spectrum and hardware efficiency, and to achieve integration gains, integrated sensing and communication (ISAC) combines sensing with communication, realizing a truly collaborative design of communication and sensing functions. ISAC possesses the capability to provide high-quality wireless communication services to target users while simultaneously offering high-precision sensing services \cite{9540344,9585321,9737357}. Given the significant advantages of ISAC, it is poised to facilitate numerous emerging applications and is widely regarded as one of the key technologies for next-generation mobile communication networks.

In recent years, the investigation of ISAC utilizing multiple-input multiple-output (MIMO) technology has attracted considerable interest. This is because MIMO can profoundly enhance system performance by deeply exploiting spatial dimension resources with beamforming design \cite{9585321,7080890}. Specifically, MIMO-based ISAC systems, equipped with multiple antennas at both the transmitter and receiver, can further increase spatial degrees of freedom {(DoF)}. Simultaneously, the system employs beamforming to concentrate signal energy in the direction of the target transmission, thereby improving the quality of ISAC signals and reducing interference to other users. Moreover, multiple data streams can be transmitted by MIMO systems within the same time-frequency resource block, which significantly improves spectrum efficiency compared to single-antenna systems \cite{1266912,1203154}. The academic community conducted a series of studies on MIMO-based ISAC systems \cite{10938928,2024arXiv240603737S,10557534,10097000,10770161,9968163,9645576}.\cite{10938928} achieved maximization of the communication rate by conducting power allocation under the condition of ensuring sensing performance. In \cite{2024arXiv240603737S}, a hybrid beamforming design was proposed by the authors to maximize the system's energy efficiency. \cite{10770161} and \cite{9968163} focused on improving the signal-to-interference-plus-noise ratio (SINR) and the eavesdropping signal-to-noise ratio using beamforming in MIMO-based ISAC systems. However, \cite{2024arXiv240603737S,10557534,10097000,10770161,9968163} did not focus on system's data transmission mode, whereas \cite{9645576} specifically studied a full-duplex data transmission model. It proposed a sensing-assisted uplink communication framework between a single-antenna user and a full-duplex base station (BS), which improved the secrecy rate by jointly optimizing radar waveforms and receiving beamforming vectors. The full-duplex model does not require different frequencies to be assigned to transmit and receive, further improving spectrum utilization compared to simplex and half-duplex systems. Although effective beamforming can improve the performance of MIMO-based ISAC systems, traditional MIMO systems face challenges due to the limitation of antennas being deployed in fixed positions. The ability to dynamically adjust the channel is limited, and when massive MIMO is deployed to improve wireless channel capacity, more antennas and radio frequency chains are required \cite{9144301}.

Fortunately, movable antenna (MA) is proposed as a promising technology to address the aforementioned challenges \cite{10972180}. In the novel MIMO systems enabled by MA, the positions of the MA elements can be dynamically adjusted in real-time by controllers, such as stepper motors or servo systems \cite{1353484,8060521,11007274}, thus fully utilizing the available {DoF} with only a limited number of antenna elements. {Recent research \cite{11224420} leverages a high-precision linear stage (positioning accuracy 0.05 mm, commensurate with sub-millimeter-wave wavelengths) integrated with closed-loop feedback control, experimentally validating mechanical actuation with measured receive-power gains exceeding 40 dB at 3.5 GHz and 23 dB at 27.5 GHz, which validate the great potential of MA technology in
improving wireless communication performance.} In the literature, various efforts have been made to further explore the potential of MA-enabled MIMO systems, validated their advantages in terms of improved system performance compared with existing systems to fixed antenna positions \cite{10684758,2024arXiv240710393D,10243545,10437926}. Both \cite{10684758} and \cite{2024arXiv240710393D} considered the positions of the MA elements as one of the optimization variables to enhance the system's security performance. \cite{10243545} investigated the joint optimization of the positions of transmitting and receiving MA elements along with the covariance matrix of the transmitted signals to maximize the capacity of MIMO systems supported by point-topology. Different from \cite{10684758,2024arXiv240710393D,10243545}, \cite{10437926} characterized the movement of MA elements as discrete motion and concurrently optimized the transmit beamforming along with the MA positioning at the BS to minimize total transmit power in multiple-input single-output (MISO) systems. Moreover, \cite{10286328} provided a general comparison with traditional fixed phased arrays (FPAs), and summarized that MA-enabled communication systems can fully leverage the spatial variations of wireless channels in limited areas, thereby increasing signal power, suppressing interference, achieving beamforming and enhancing spatial multiplexing performance. Overall, MA-enabled novel MIMO systems can significantly enhance system performance by leveraging the characteristics of MA.

Owing to its numerous advantages, MA has garnered extensive attention from both industry and academia \cite{10286328,10278220,10654366,10694747,2024arXiv240320025D,2024arXiv240703228W,10946353,ma2025movableantennaenhancedintegrated,10696953}. The application of MA research in industrial Internet-of-things (IoT), satellite communication, smart homes, and other fields is expected to further promote industry development \cite{10286328}. Current research not only focuses on the comparison between traditional MIMO systems and MA-enabled MIMO systems but also extends to various communication system models enabled by MA. \cite{10278220} studied the enhanced beamforming of MA arrays by utilizing new DoF through antenna position optimization, allowing for interference mitigation between communication and sensing signals as well as multiuser interference for communication.\cite{10654366} investigated the communication quality of multi-user in a unmanned aerial vehicle (UAV) system enhanced by the MA. \cite{10694747} studied a system featuring multiple MAs at the BS and multiple sets of users each equipped with single MA, with the objective of maximizing the minimum weighted signal-to-noise ratio across all users. \cite{2024arXiv240320025D} examined the physical layer security of a MA-enabled full-duplex system. Due to the mobility of the elements of the MA, the integration of MA and ISAC systems can further leverage the characteristics of wireless channel spatial variation to enhance system performance. It can dynamically and precisely adjust antenna element positions based on real-time sensing and communication needs. By adjusting MA elements to suitable positions, the system can fully exploit channel gain characteristics, enhancing signal transmission quality, reducing the bit-error rate, and significantly improving communication performance. It also enables more accurate capture of target signal characteristics, improves parameter estimation accuracy like target position and velocity, and boosts sensing capabilities. In particular, \cite{2024arXiv240703228W} showed the similarity of user channels in ISAC systems can be reduced by applying MA, thereby improving channel gain. \cite{10946353} significantly improved the communication rate and sensing mutual information of ISAC compared to fixed uniform arrays. \cite{ma2025movableantennaenhancedintegrated} enhanced communication and sensing performance of the ISAC system by optimizing antenna positions and designed an alternating optimization algorithm to maximize the minimum achievable rate for communication users and minimize sensing angle estimation error. In addition, \cite{10696953} explored the potential of MA in enhancing the performance of ISAC. However, research on MA-enabled ISAC systems is relatively scarce and is in its infancy.

Based on the aforementioned discussion, this paper primarily considers a minimization of transmit power consumption for the MA-enabled full-duplex ISAC system. Specifically, we consider a more practical case where MA has a spatially discrete set of candidate positions. {The discrete movement is motivated by the need to simplify the control and reduce the complexity of the mechanical systems \cite{8060521,7360379}. In contrast to the continuous positions of the MA system, we consider a more practical model, which is more amenable to deployment in practical applications.} The objective is to minimize the total transmit power consumption through the joint optimization of the MA elements' discrete positions, beamforming vectors, sensing signal covariance matrix and user transmit power. Given the highly coupled optimization variables, we aim to design an efficient joint optimization algorithm for the MA-enabled full-duplex ISAC system to address the transmit power consumption minimization problem. The main contributions of this paper are summarized as follows:
\begin{itemize}
\item{We propose a novel MA-enabled full-duplex ISAC system, where the deployed MA can improve the system performance by increasing the spatial {DoF}. In this paper, we adopt a more practical MA model, where the candidate positions of MA elements are discrete. Then, a total transmit power consumption minimization problem is formulated via jointly optimizing the discrete position of MA elements, beamforming vectors, sensing signal covariance matrix and user transmit power. Considering the interdependence of the optimization variables, the minimization problem is non-convex, rendering it difficult to attain solutions.}
\item{To address the issue of minimizing the total transmit power consumption of the system, we propose a discrete binary particle swarm optimization (BPSO) algorithm framework. More specifically, by applying the difference-of-convex (DC) programming and successive convex approximation (SCA) to transform the non-convex and rank-1 terms in fitness function, we can iteratively solve the fitness function to determine the discrete positions of MA elements. Given the obtained discrete positions of MA elements, the solutions for beamforming vectors, sensing signal covariance matrix and user transmit power are also determined.}
\item{Numerical simulation results indicate that the system has performance advantages over traditional ISAC systems. Due to the enhanced spatial {DoF} that MA offers, the MA-enabled ISAC system can reduce the system's total transmit power consumption compared with fixed antenna arrays. In addition, beampattern simulation also shows that the proposed BPSO-based optimization algorithm framework in the MA-enabled full-duplex ISAC network can achieve multi-beam alignment and interference suppression to a certain extent.}
\end{itemize}

The structure of this paper is organized as follows. In Section II, we present the MA-enabled full-duplex ISAC system model and the formulation of total transmit power consumption minimization problem. Section III provides the joint discrete antenna positioning and beamforming optimization algorithm based on BPSO algorithm framework. In Section IV, numerical results elaborate that the proposed system has better performance in terms of reducing the transmit power consumption compared with fixed antenna array. Finally, Section V concludes the paper. 

\textit{Notation}: Lower-case letters are used to represent scalar values. Vectors are indicated with bold lower-case letters, while matrices are represented by bold upper-case letters. $\left(  \cdot  \right)^{T}$ and $\left(  \cdot  \right)^{H}$ respectively denote transpose and conjugate transpose of a matrix. ${\left(  \cdot  \right)^{{\text{ - }}1}}$  represents the inverse of the matrix. ${\text{Tr}}\left(  \cdot  \right)$ denotes the trace of a square matrix. Denote $\left\|  \cdot  \right\|$ by norm of a vector and $\left|  \cdot  \right|$ by the absolute value of a scalar. Let ${\text{rank}}\left(  \cdot  \right)$ and ${\left[  \cdot  \right]_{i,j}}$ return the rank and the $\left( {i,j} \right)$-th entry of a matrix, respectively. ${\mathbb{C}^{M \times N}}$ and ${\mathbb{R}^{M \times N}}$ are the sets of $M \times N$-dimensional complex and real matrices. ${{\mathbf{I}}_N}$ refers to the identity matrix of dimension $N$. ${\mathbf{A}} \succeq 0$ indicates that $\bf{A}$ is a positive semidefinite (PSD) matrix. Let ${\mathbf{A}} \succ 0$ imply that $\bf{A}$ is a positive definite matrix. We use $\mathbb{E}\left\{  \cdot  \right\}$ for the expectation operation. $j$ denotes the imaginary unit, i.e., $j^2 = -1$. $\left\langle {{\bf{A}}, {\bf{B}}} \right\rangle $ represents the Frobenius inner product of matrices $\bf{A}$ and $\bf{B}$. $\partial \left\| {\bf{A}} \right\|$ represents the subgradient of the spectral norm of the matrix $\bf{A}$. \% represents the modulo operation.

\section{MA-Enabled Full-Duplex ISAC System Model and Problem Formulation}
Firstly, we consider the model of a MA-enabled ISAC system with full-duplex operation, as illustrated in Fig. 1. Specifically, the full-duplex dual-function radar and communication base station (FD-DFRC-BS) is equipped with two MAs for receiving and transmitting signals. The receiving MA is used to receive communication signals from $U$ single-antenna uplink users and echo signals from radar sensing targets, while simultaneously being subject to $K$ non-target radar sensing interference. We assume the clutter in the surrounding environment as the interference in radar sensing. Without loss of generality, we follow the method of the interference source in \cite{6649991} for modeling. In this model, the sensing signals are reflected towards the BS, thereby producing undesired interference. The transmitting MA sents downlink ISAC signals over the same time-frequency resources, communicates with $D$ single-antenna downlink users, and senses a radar sensing target. The number of elements in the receiving MA is ${N_r}$, and the number of elements in the transmitting MA is ${N_t}$. This paper considers a more practical MA, where the candidate positions of the antenna elements are discrete. Specifically, it is assumed that all elements of the transmitting MA have a total of $N$ candidate discrete positions ($N > {N_t}$). The matrix representing these candidate discrete positions are denoted as 
${{\mathbf{P}}_t}{\text{ = }}\left[ {{{\mathbf{p}}_{t,1}}, \ldots ,{{\mathbf{p}}_{t,N}}} \right]$, where ${{\mathbf{p}}_{t,n}} = {\left[ {{x_{t,n}},{y_{t,n}}} \right]^T}$. All receiving MA elements have a total of $M$ candidate discrete positions $(M > {N_r})$, and the corresponding candidate discrete position matrix is expressed as ${{\mathbf{P}}_r}{\text{ = }}\left[ {{{\mathbf{p}}_{r,1}}, \ldots ,{{\mathbf{p}}_{r,M}}} \right]$, where ${{\mathbf{p}}_{r,M}} = {\left[ {{x_{r,m}},{y_{r,m}}} \right]^T}$. Note that in this paper, the distance between candidate discrete positions of the MA is equal to $l$ in either the horizontal or vertical direction. Additionally, the full-duplex mode allows for simultaneous operation in both directions of communication (transmitting and receiving) without mutual interference. {Theoretically}, for communication, this improves the efficiency of data transmit and the utilization of bandwidth, while for radar sensing, the available bandwidth increases, enhancing the radar's sensing performance. Moreover, the MA leverages the flexibility of moving MA elements to exploit the {DoF} available in the area, and the discretization of MA element candidate positions facilitates their deployment and application in practical systems. Meanwhile, for the sake of analysis, this paper assumes that channel state information (CSI) can be perfectly obtained at the FD-DFRC-BS \cite{10978584,wang2024movableantennaschannelmeasurement}. The scenario of more practical imperfect CSI can be considered as future research work.
\begin{figure}[!t]
\centering
\includegraphics[width=3.3in]{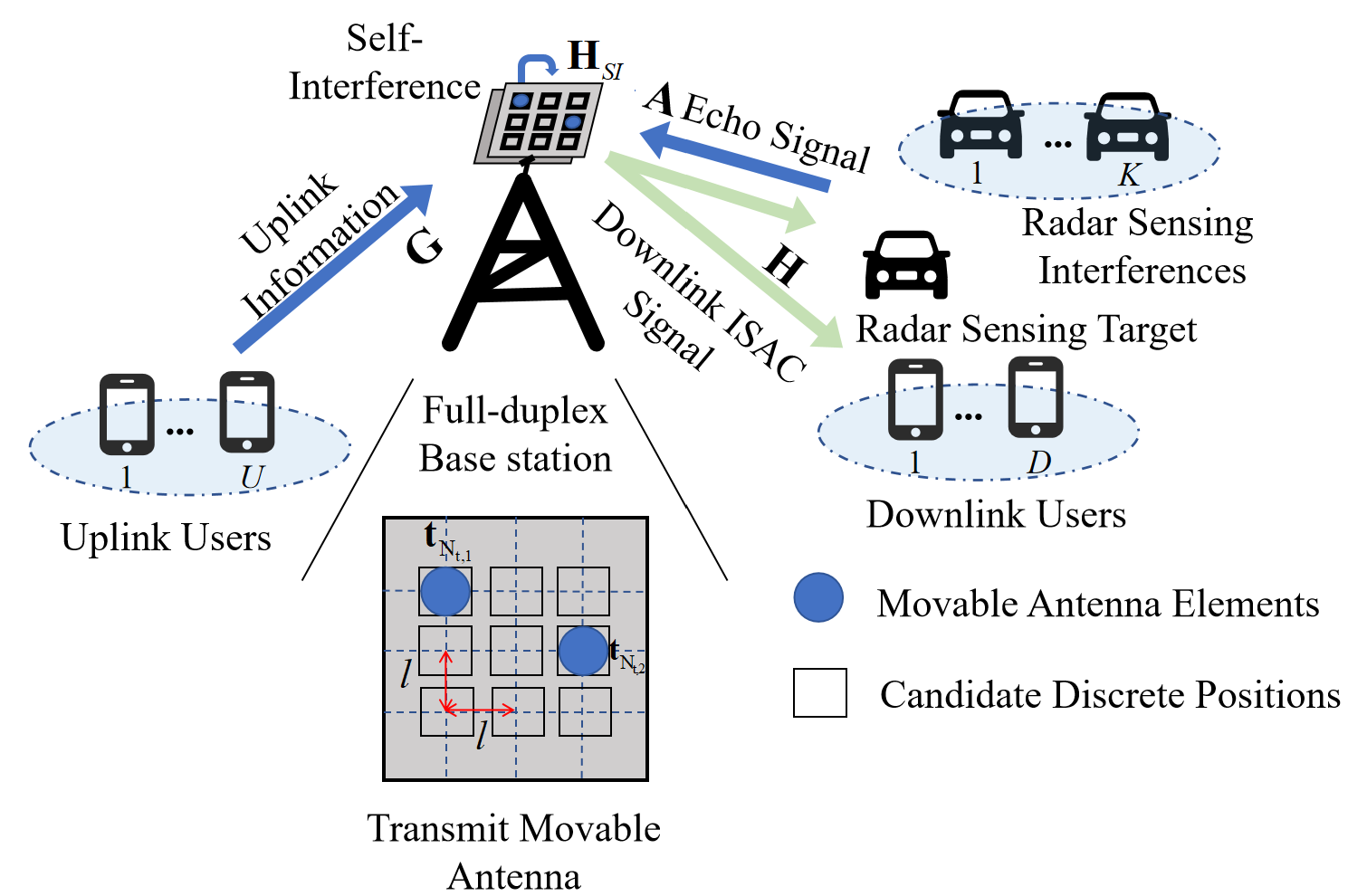}
\caption{MA-enabled full-duplex ISAC system.}
\label{fig1}
\end{figure}
\subsection{Channel Model}
In this subsection, we primarily elaborate on the channel model for radar sensing and communication in the MA-enabled full-duplex ISAC system. Firstly, we introduce the radar sensing channel model of the system. Specifically, for the MA system, its MA elements can move among the candidate discrete positions within a two-dimensional area, leading to changes in the physical channel. Assuming that the channel sensing information is line-of-sight (LoS), taking the receiving MA as an example, the actual location vector of the ${n_r}$-th MA element is expressed as ${{\mathbf{t}}_{r{\text{,}}{n_r}}} = {\left[ {{t_{{r_x},{n_r}}},{t_{{r_y},{n_r}}}} \right]^T}$, which can be represented by a two-dimensional vector ${{\mathbf{b}}_{r,{n_r}}}$ as follows: ${{\mathbf{t}}_{r{\text{,}}{n_r}}}{\text{ = }}{{\mathbf{P}}_r}{{\mathbf{b}}_{r,{n_r}}}$, where ${{\mathbf{b}}_{r,{n_r}}} = {\left[ {{b_{r,{n_r}}}\left[ 1 \right],...,{b_{r,{n_r}}}\left[ M \right]} \right]^T}$, $\forall {n_r} \in \left\{ {1,...,{N_r}} \right\}$, ${b_{r,{n_r}}}\left[ m \right] \in \left\{ {0,1} \right\}$, $\forall m \in \left\{ {1,...,M} \right\}$ and $\sum\nolimits_{m = 1}^M {{b_{r,{n_r}}}\left[ m \right] = 1}$. Similarly, the actual position of the ${n_t}$-th transmitting MA element is defined as ${{\mathbf{t}}_{t{\text{,}}{n_t}}} = {\left[ {{t_{{t_x},{n_t}}},{t_{{t_y},{n_t}}}} \right]^T}$, and a binary direction vector 
${{\mathbf{b}}_{t,{n_t}}}$ is used to express it as 
${{\mathbf{t}}_{t,{n_t}}}{\text{ = }}{{\mathbf{P}}_t}{{\mathbf{b}}_{t,{n_t}}}$. The binary direction  ${{\mathbf{b}}_{t,{n_t}}}$ is given by ${{\mathbf{b}}_{t,{n_t}}} = {\left[ {{b_{t,{n_t}}}\left[ 1 \right],...,{b_{t,{n_t}}}\left[ N \right]} \right]^T}$, $\forall {n_t} \in \left\{ {1,...,{N_t}} \right\}$ and it follows that ${b_{t,{n_t}}}\left[ n \right] \in \left\{ {0,1} \right\}$, $\forall n \in \left\{ {1,...,N} \right\}$ and $\sum\nolimits_{n = 1}^N {{b_{t,{n_t}}}\left[ n \right] = 1}$. Let the angles be the elevation angle and azimuth angle between the signal and the radar sensing target $\theta$ , $\phi  \in \left[ {{{ - \pi } \mathord{\left/
 {\vphantom {{ - \pi } 2}} \right.
 \kern-\nulldelimiterspace} 2},{{ - \pi } \mathord{\left/
 {\vphantom {{ - \pi } 2}} \right.
 \kern-\nulldelimiterspace} 2}} \right]$, respectively. Therefore, the steering vector for the transmitting MA array is expressed as
 \begin{equation}
     \begin{aligned}
        {{\mathbf{a}}_t}\left( {\theta ,\phi } \right){\text{ = }}\frac{1}{{\sqrt {{N_t}} }}\left[ {1,{e^{\frac{{j2\pi \left( {\left( {{t_{{t_x},2}} - {t_{{t_x},1}}} \right)\cos \theta \sin \phi  + ({t_{{t_y},2}} - {t_{{t_y},1}})\sin \theta } \right)}}{\lambda }}},}\right.\\ \left.{\ldots ,{e^{\frac{{j2\pi \left( {\left( {{t_{{t_x},{N_t}}} - {t_{{t_x},1}}} \right)\cos \theta \sin \phi  + \left( {{t_{{t_y},{N_t}}} - {t_{{{\text{t}}_y},1}}} \right)\sin \theta } \right)}}{\lambda }}}} \right]^T.
     \end{aligned}
 \end{equation}
 Similarly, the steering vector for the receiving MA array is expressed as 
 \begin{equation}
     \begin{aligned}
        {{\mathbf{a}}_r}\left( {\theta ,\phi } \right){\text{ = }}\frac{1}{{\sqrt {{N_r}} }}\left[ {1,{e^{\frac{{j2\pi \left( {\left( {{t_{{r_x},2}} - {t_{{r_x},1}}} \right)\cos \theta \sin \phi  + ({t_{{r_y},2}} - {t_{{r_y},1}})\sin \theta } \right)}}{\lambda }}}, }\right.\\ \left.{\ldots ,{e^{\frac{{j2\pi \left( {\left( {{t_{{r_x},{N_t}}} - {t_{{r_x},1}}} \right)\cos \theta \sin \phi  + ({t_{{r_y},{N_r}}} - {t_{{r_y},1}})\sin \theta } \right)}}{\lambda }}}} \right]^T.
     \end{aligned}
 \end{equation}

 Then, we describe the communication channel model of the system. Based on the field response-based channel model, the channel between the transmitting and receiving MA can be modeled as ${{\mathbf{H}}_{{\text{SI}}}} \in {\mathbb{C}^{{N_r} \times {N_t}}}$, the channel between $U$ uplink users and the receiving MA can be modeled as ${\mathbf{G}} \in {\mathbb{C}^{U \times {N_r}}}$, and the channel between $D$ downlink users and the transmitting MA can be modeled as ${\mathbf{H}} \in {\mathbb{C}^{D \times {N_t}}}$ as follows.

For the FD-DFRC-BS, the channel between the transmitting and receiving MA is denoted as ${{\mathbf{H}}_{{\text{SI}}}}$, where the channel from the ${n_r}$-th receiving MA element to the ${n_t}$-th transmitting MA element is expressed as ${\left[{{\mathbf{H}}_{{\text{SI}}}} \right]_{{n_r},{n_t}}} = \sqrt {\eta _{{n_r},{n_t}}^{\text{SI}}} {e^{ - j2\pi \frac{{{d_{{n_r},{n_t}}}}}{\lambda }}}$, which represents the element in the ${n_r}$-th row and ${n_t}$-th column of matrix ${{\mathbf{H}}_{{\text{SI}}}}$. Here, $\eta _{{n_r},{n_t}}^\text{{SI}}$ and ${d_{{n_r},{n_t}}}$ denote the path loss and distance between the ${n_r}$-th receiving MA element and the ${n_t}$-th transmitting MA element.

For $U$ users in the uplink and the receiving MA, the channel matrix $\bf{G}$ is expressed as ${\mathbf{G}}{\text{ = }}\left[ {{{\mathbf{g}}_1}\left( {{{\mathbf{t}}_{r{\text{,1}}}}} \right){\text{,}}...{\text{,}}{{\mathbf{g}}_{{N_r}}}\left( {{{\mathbf{t}}_{r{\text{,}}{N_r}}}} \right)} \right] \in {\mathbb{C}^{U \times {N_r}}}$, where ${{\mathbf{g}}_{{n_r}}}\left( {{{\mathbf{t}}_{r{\text{,}}{n_r}}}} \right)$ represents the uplink channel between $U$ users and the ${n_r}$-th MA element. It is expressed as ${{\mathbf{g}}_{{n_r}}}\left( {{{\mathbf{t}}_{r{\text{,}}{n_r}}}} \right) = {\left[ {{g_{{n_r}}}_{,1}\left( {{{\mathbf{t}}_{r{\text{,}}{n_r}}}} \right),...,{g_{{n_r}}}{{_,}_U}\left( {{{\mathbf{t}}_{r{\text{,}}{n_r}}}} \right)} \right]^T} \in {\mathbb{C}^{U \times 1}}$, where ${g_{{n_r}}}_{,u}\left( {{{\mathbf{t}}_{r{\text{,}}{n_r}}}} \right) \in \mathbb{C}$ denotes the channel coefficient between the
${n_r}$-th receiving MA element and the $u$-th uplink user. To facilitate subsequent processing, we define the channel between the $U$ uplink users and the $M$ candidate discrete positions of the receiving MA as ${\mathbf{\hat G}}{\text{ = }}\left[ {{{{\mathbf{\hat G}}}_1},...,{{{\mathbf{\hat G}}}_{{N_r}}}} \right] \in {\mathbb{C}^{U \times M{N_r}}}$, where $
{{\mathbf{\hat G}}_{{n_r}}}{\text{ = }}\left[ {{{\mathbf{g}}_{{n_r}}}\left( {{{\mathbf{p}}_{r,1}}} \right),...,{{\mathbf{g}}_{{n_r}}}\left( {{{\mathbf{p}}_{r,M}}} \right)} \right] \in {\mathbb{C}^{U \times M}}$ represents the channel vector of the $M$ candidate discrete positions from $U$ uplink users to the ${n_r}$-th receiving MA element. The channel vector at the candidate discrete position 
${{\mathbf{p}}_{r,m}}$ for the $U$ uplink users and the 
${n_r}$-th receiving MA element is given by ${{\mathbf{g}}_{{n_r}}}\left( {{{\mathbf{p}}_{r,m}}} \right) = {\left[ {{g_{{n_r}}}_{,1}\left( {{{\mathbf{p}}_{r{\text{,}}m}}} \right),...,{g_{{n_r}}}{{_,}_U}\left( {{{\mathbf{p}}_{r{\text{,}}m}}} \right)} \right]^T} \in {\mathbb{C}^{U \times 1}}$. Here, ${g_{{n_r}}}_{,u}\left( {{{\mathbf{p}}_{r{\text{,}}m}}} \right) \in \mathbb{C}$ denotes the channel coefficient between the ${n_r}$-th receiving MA element and the $u$-th uplink user, when the latter is located at the candidate discrete position ${{\mathbf{p}}_{r,m}}$. Suppose that
${l_p} \in \left\{ {1,...,{L_p}} \right\}$ represents the number of paths from the ${n_r}$-th receiving MA element to the $u$-th uplink user, let $\theta _{_{u,{l_p}}}^U$ and $\phi _{_{u,{l_p}}}^U$ respectively represent the elevation angle and azimuth angle of the ${l_p}$-th path of the $u$-th uplink user, then ${g_{{n_r}}}_{,u}\left( {{{\mathbf{p}}_{{N_r}{\text{,}}m}}} \right)$ can be expressed as 
\begin{equation}
\begin{aligned}
    {g_{{n_r}}}_{,u}\!\left(\! {{{\bf{p}}_{{N_r}{\rm{,}}m}}} \!\right)\!\! =\!\! \sqrt {{\alpha _{u,{l_1}}}}\! {e^{\frac{{j2\pi \left(\! {\left( {{x_{r,m}}\! -\! {x_{r,1}}} \right)\!\cos \theta _{u,{l_1}}^U\!\!\!\sin\! \phi _{u,{l_1}}^U \!\!\!+\! \left( {{y_{r,m}} \!-\! {y_{r,1}}}\! \right)\!\sin \theta _{u,{l_1}}^U}\!\! \right)}}{\lambda }}}\\ + ... + \sqrt {{\alpha _{u,{l_p}}}} {e^{\frac{{j2\pi \left( {\left( {{x_{r,m}} - {x_{r,1}}} \right)\cos \theta _{u,{l_p}}^U\sin \phi _{u,{l_p}}^U + \left( {{y_{r,m}} - {y_{r,1}}} \right)\sin \theta _{u,{l_p}}^U} \right)}}{\lambda }}},
\end{aligned}
\end{equation}where $\alpha _{u,{l_p}}$ is the channel fading factor of the $l_p$ path from the FD-DFRC-BS to the $u$-th uplink user. Due to the discretization of the MA positions, where the center-to-center distance between adjacent candidate discrete positions is half a wavelength, we carried out the modeling with reference to the planar antenna array. For the channel modeling of MA with continuous candidate positions, please refer specifically to \cite{10318061}. Let ${{\mathbf{\hat g}}_u}$ denote the $u$-th row of ${\mathbf{\hat G}}$, which can represent the channel of the $u$-th uplink user and the $M$ candidate discrete positions of the ${N_r}$ receiving MA elements. The channel matrix $\bf{G}$ between $U$ uplink users and receiving MA can be represented as 
\begin{equation}
    {\mathbf{G}}{\text{ = }}{\mathbf{\hat G}}{{\mathbf{B}}_r},
\end{equation}
where ${{\mathbf{B}}_r} \in {\mathbb{C}^{{N_r}M \times {N_r}}}$. It can also be specifically expressed as
\begin{equation}
{{\mathbf{B}}_r}{\text{ = }}\left[ \begin{matrix}
  {{\mathbf{b}}_{r,}}_1   & {{\mathbf{0}}_M}   &  \cdots &     {{\mathbf{0}}_M}  \hfill \\
  {{\mathbf{0}}_M}  &   {{\mathbf{b}}_{r,2}}  &  \cdots &    {{\mathbf{0}}_M} \hfill \\
   \cdots    &     \cdots   &    \cdots    &  \cdots  \hfill \\
  {{\mathbf{0}}_M}   &  {{\mathbf{0}}_M}    & \cdots  &   {{\mathbf{b}}_{r,{N_r}}} \hfill \\ 
\end{matrix}  \right].
\end{equation}

In addition, for the wireless channel of this system between $D$ downlink users and the transmitting MA, we have
${\mathbf{H}}{\text{ = }}\left[ {{{\mathbf{h}}_1}\left( {{{\mathbf{t}}_{t{\text{,1}}}}} \right){\text{,}}...{\text{,}}{{\mathbf{h}}_{{N_t}}}\left( {{{\mathbf{t}}_{t{\text{,}}{N_t}}}} \right)} \right] \in {\mathbb{C}^{D \times {N_t}}},$ where 
${{\mathbf{h}}_{{n_t}}}\left( {{{\mathbf{t}}_{{\text{t,}}{n_t}}}} \right)$ represents the channel coefficient between the ${n_t}$-th MA element and \textit{D} downlink user. It can be expressed as %
${{\mathbf{h}}_{{n_t}}}\left( {{{\mathbf{t}}_{t{\text{,}}{n_t}}}} \right) = {\left[ {{h_{{n_t}}}_{,1}\left( {{{\mathbf{t}}_{t{\text{,}}{n_t}}}} \right),...,{h_{{n_t}}}{{_,}_D}\left( {{{\mathbf{t}}_{t{\text{,}}{n_t}}}} \right)} \right]^T} \in {\mathbb{C}^{D \times 1}}$, where ${h_{{n_t}}}_{,d}\left( {{{\mathbf{t}}_{t{\text{,}}{n_t}}}} \right) \in \mathbb{C}$ represents the channel coefficient between the ${n_t}$-th transmitting MA element and the $d$-th downlink user. In order to facilitate subsequent processing, we define the channel between $D$ different users and $N$ candidate discrete positions of transmitting MA as
${\mathbf{\hat H}}{\text{ = }}\left[ {{{{\mathbf{\hat H}}}_1},...,{{{\mathbf{\hat H}}}_{{N_t}}}} \right] \in {\mathbb{C}^{D \times N{N_t}}}$. ${{\mathbf{\hat H}}_{{n_t}}}{\text{ = }}\left[ {{{\mathbf{h}}_{{n_t}}}\left( {{{\mathbf{p}}_{t,1}}} \right),...,{{\mathbf{h}}_{{n_t}}}\left( {{{\mathbf{p}}_{t,N}}} \right)} \right] \in {\mathbb{C}^{D \times N}}$ is the channel matrix from the $D$ downlink users to the $N$ candidate discrete channels of the ${n_t}$-th transmitting MA element. Additionally, 
${{\mathbf{h}}_{{n_t}}}\left( {{{\mathbf{p}}_{t,n}}} \right) = {\left[ {{h_{{n_t}}}_{,1}\left( {{{\mathbf{p}}_{t,n}}} \right),...,{h_{{n_t}}}{{_,}_D}\left( {{{\mathbf{p}}_{t,n}}} \right)} \right]^T} \in {\mathbb{C}^{D \times 1}}$ represents the channel vector of $D$ downlink users and the ${n_t}$-th transmitting MA element at the candidate discrete position ${{\mathbf{p}}_{t,n}}$. Here, ${h_{{n_t}}}_{,d}\left( {{{\mathbf{p}}_{t{\text{,}}n}}} \right) \in \mathbb{C}$ represents the channel coefficient between the ${n_t}$-th transmitting MA element and the $d$-th downlink user. Let $\forall{l_p} \in \left\{ {1,...,{L_p}} \right\}$ represent the number of paths between the ${n_t}$-th transmitting MA element and the $d$-th downlink user, where are a total of ${L_p}$ such paths. Denote by $\theta _{_{d,{l_p}}}^D$, $\phi _{_{d,{l_p}}}^D$ the elevation angle and azimuth angle for the ${l_p}$-th path to the $d$-th downlink user, respectively, then ${h_{{n_t}}}_{,d}\left( {{{\mathbf{p}}_{{N_t}{\text{,}}n}}} \right)$ can be expressed as
\begin{equation}
    \begin{aligned}
        {h_{{n_t}}}_{,d}\!\left(\! {{{\bf{p}}_{{N_t}{\rm{,n}}}}}\! \right)\!\! = \!\!\sqrt {{\alpha _{d,{l_1}}}} {e^{\frac{{j2\pi \left( \!{\left( {{x_{t,n}}\!\! -\! {x_{t,1}}} \right)\cos \theta _{d,{l_1}}^D\!\!\sin \phi _{d,{l_1}}^D\!\! +\! \left( {{y_{t,n}}\!\! -\! {y_{t,1}}} \right)\sin \theta _{d,{l_1}}^D}\! \right)}}{\lambda }}} \\+\! ...\! + \sqrt {{\alpha _{d,{l_p}}}} {e^{\frac{{j2\pi \left( {\left( {{x_{t,n}} \!-\! {x_{t,1}}} \right)\cos \theta _{d,{l_p}}^D\!\!\sin \phi _{d,{l_p}}^D + \left( {{y_{t,n}}\! -\! {y_{t,1}}} \right)\sin \theta _{d,{l_p}}^D} \right)}}{\lambda }}},
    \end{aligned}
\end{equation}where $\alpha _{d,{l_p}}$ is the channel fading factor of the $l_p$ path from the FD-DFRC-BS to the $d$-th downlink user. Let ${{\mathbf{\hat h}}_d}$ denote the $d$-th row of ${\mathbf{\hat H}}$, which represents the channel gain between the $d$-th downlink user and the $N$ candidate discrete positions of the ${N_t}$ transmitting MA elements. The channel matrix $\bf{H}$ between $D$ downlink users and transmitting MA can be represented as 
\begin{equation}
    {\mathbf{H}}{\text{ = }}{\mathbf{\hat H}}{{\mathbf{B}}_t},
\end{equation}
where ${{\mathbf{B}}_t} \in {\mathbb{C}^{{N_t}N \times {N_t}}}$, specifically represented as
\begin{equation}
{{\mathbf{B}}_t}{\text{ = }}\left[ \begin{matrix}
  {{\mathbf{b}}_{t,}}_1  & {{\mathbf{0}}_N}   &  \cdots &    {{\mathbf{0}}_N}  \hfill \\
  {{\mathbf{0}}_N}   & {{\mathbf{b}}_{t,2}}   &  \cdots   &  {{\mathbf{0}}_N} \hfill \\
   \cdots    &   \cdots   &     \cdots  &    \cdots  \hfill \\
  {{\mathbf{0}}_N}  &  {{\mathbf{0}}_N}   &  \cdots  &   {{\mathbf{b}}_{t,{N_t}}} \hfill \\ 
\end{matrix}  \right].
\end{equation}

\subsection{Signal Model}
For the downlink transmission process of the MA-enabled full-duplex ISAC system, the FD-DFRC-BS transmits narrowband ISAC signals ${\mathbf{x}} \in {\mathbb{C}^{{N_t} \times 1}}$ via multi-antenna beamforming, which serves for radar target sensing and downlink communication for multiple users. According to \cite{9916163,9652071}, it can be represented as 
\begin{equation}
    {\mathbf{x}} = \sum\limits_{d = 1}^D {{{\mathbf{w}}_d}{s_d} + {{\mathbf{s}}_0}},
\end{equation}
where ${{\mathbf{w}}_d} \in {\mathbb{C}^{{N_t} \times 1}}$ represents the beamforming vector associated with the $d$ downlink user. ${s_d}$ denotes the data transmitted to the $d$-th user for communication, satisfying $\mathbb{E}\left\{ {{{\left| {{s_d}} \right|}^2}} \right\} = 1$. ${{\mathbf{s}}_0} \in {\mathbb{C}^{{N_t} \times 1}}$ represents the dedicated sensing signal sent to the radar sensing target, determined by ${{\mathbf{W}}_0} \triangleq \mathbb{E}\left\{ {{\mathbf{s}}_0}{{\mathbf{s}}_0^H} \right\}$ and is mutually orthogonal to $\left\{ {{s_d}} \right\}_{d = 1}^D$. Therefore, the signal received by the $d$-th user in the downlink can be expressed as 
\begin{equation}
    \begin{aligned}
        y_d^{{\text{DL}}} &= \underbrace {{{ { {{{\mathbf{\hat h}}}_d}{{\mathbf{B}}_t}} }}{{\mathbf{w}}_d}{s_d}}_{{\text{Desired signal}}} + \underbrace {\sum\limits_{d' \ne d}^D {{{ { {{{\mathbf{\hat h}}}_d}{{\mathbf{B}}_t}} }}{{\mathbf{w}}_{d'}}{s_{d'}}} }_{{\text{Multiuser interference}}} 
        + \underbrace {{{ { {{{\mathbf{\hat h}}}_d}{{\mathbf{B}}_t}} }}{{\mathbf{s}}_0}}_{{\text{Sensing signal}}} + {n_d},\forall d,
    \end{aligned}
\end{equation}where ${n_d}$ represents the additive Gaussian white noise (AWGN) with a variance of $\sigma _d^2$ introduced at the receiving end of the $d$-th downlink user.

Then, considering the uplink transmission process of the system, when FD-DFRC-BS communicates and senses in the downlink, it will receive the data sent by the uplink users as well as the sensing echo signal from the radar. Suppose that the $u$-th uplink user sends data ${d_u} \in \mathbb{C}$. Supposed to be detected sensing target is located at ${\theta _0},{\phi _0}$, then the reflection of target is ${\beta _0}{{\mathbf{a}}_r}\left( {{\theta _0},{\phi _0}} \right){\mathbf{a}}_t^H\left( {{\theta _0}, {\phi _0}} \right){\mathbf{x}}$, where ${\beta _0} \in \mathbb{C}$ represents the gain of the target sensing channel, mainly influenced by path loss and radar. Using parameter estimation scheme \cite{9724174,9537599}, we can obtain parameters ${\beta _0}$, ${\theta _0}$, ${\phi _0}$ at FD-DFRC-BS, so the received signal at FD-DFRC-BS can be expressed as 
    \begin{equation}
    \begin{aligned}
        {{\mathbf{y}}^{{\text{BS}}}} &= \underbrace {\sum\limits_{u = 1}^U {\left({{{\mathbf{\hat g}}}_u}{{\mathbf{B}}_r}\right)}^H {d_u}}_{{\text{Communication signal}}} + \underbrace {{\beta _0}{\mathbf{A}}\left( {{\theta _0},{\phi _0}} \right){\mathbf{x}}}_{{\text{Target reflection}}}\\& + \underbrace {\sum\limits_{k = 1}^K {{\beta _k}} {\mathbf{{A}}}\left( {{\theta _k},{\phi _k}} \right){\mathbf{x}}}_{{\text{Non-target reflection}}} + \underbrace {{{\mathbf{H}}_{{\text{SI}}}}{\mathbf{x}}}_{{\text{SI}}} + {\mathbf{n}},
    \end{aligned}
    \end{equation}
where ${\mathbf{A}}\left( {\theta ,\phi } \right) = {{\mathbf{a}}_r}\left( {\theta ,\phi } \right){\mathbf{a}}_t^H\left( {\theta ,\phi } \right)$, ${\alpha _u}$ is the channel attenuation of the $u$-th uplink user to FD-DFRC-BS, ${\mathbf{n}} \in {\mathbb{C}^{{N_r} \times 1}}$ represents the AWGN with variance $\sigma _r^2{{\bf{I}}_{{N_r}}}$. ${\beta _k}$ is the amplitude of the interference signal introduced by the $k$-th non-target sensing radar. 

\newcounter{my1}
\begin{figure*}[!t]
	\normalsize
	\setcounter{my1}{\value{equation}}
	\setcounter{equation}{11}
	\begin{equation}
		  \begin{aligned}
		          {\text{SIN}}{{\text{R}}^r} &= \frac{{\mathbb{E}\left\{ {|{\beta _0}{{\mathbf{v}}^H}{\mathbf{A}}\left( {{\theta _0},{\phi _0}} \right){\mathbf{x}}{|^2}} \right\}}}{{\sum\limits_{u = 1}^U {\mathbb{E}\left\{ {{{\left| \left({{\mathbf{v}}} {{{\mathbf{\hat g}}}_u}{{\mathbf{B}}_r}\right)^H{d_u}\right|}^2}} \right\} + \sum\limits_{k = 1}^K {\mathbb{E}\left\{ {{{\left| {{{\mathbf{v}}^H}({\beta _k}{\mathbf{A}}({\theta _k},{\phi _k}) + {{\mathbf{H}}_{{\text{SI}}}}){\mathbf{x}}} \right|}^2}} \right\} + \mathbb{E}\left\{ {{{\left| {{{\mathbf{v}}^H}{\mathbf{n}}} \right|}^2}} \right\}} } }}\\& = \frac{{{{\left| {{\beta _0}} \right|}^2}{{\bf{v}}^H}{\bf{A}}\left( {{\theta _0},{\phi _0}} \right)\left( {\sum\limits_{d = 1}^D {{{\bf{w}}_d}{\bf{w}}_d^H + {{\bf{W}}_0}} } \right){\bf{A}}{{\left( {{\theta _0},{\phi _0}} \right)}^H}{\bf{v}}}}{{{{\bf{v}}^H}\left( {\sum\limits_{u = 1}^U {{p_u}{{\left( {{{{\bf{\hat g}}}_u}{{\bf{B}}_r}} \right)}^H{{{\bf{\hat g}}}_u}{{\bf{B}}_r}}}  + {\bf{Q}}\left( {\sum\limits_{d = 1}^D {{{\bf{w}}_d}{\bf{w}}_d^H + {{\bf{W}}_0}} } \right){{\bf{Q}}^H} + \sigma _r^2{{\bf{I}}_{{N_r}}}} \right){\bf{v}}}},
		  \end{aligned}
	\end{equation}

        \begin{equation}
            {\text{SINR}}_u^U = \frac{{{p_u}\left({{\bf{r}}_u {{{{\bf{\hat g}}}_u}{{\bf{B}}_r}} }\right)^H{{{\bf{\hat g}}}_u}{{\bf{B}}_r}{{\bf{r}}_u}}}{{{\bf{r}}_u^H\left( {\sum\limits_{u' \ne u}^U {{p_{u'}}{{\left( {{{{\bf{\hat g}}}_{u'}}{{\bf{B}}_r}} \right)}^H}{{{\bf{\hat g}}}_{u'}}{{\bf{B}}_r} + {\bf{C}}\left( {\sum\limits_{d = 1}^D {{{\bf{w}}_d}{\bf{w}}_d^H + {{\bf{W}}_0}} } \right){{\bf{C}}^H} + \sigma _r^2{{\bf{I}}_{{N_r}}}} } \right){{\bf{r}}_u}}},\forall u,
        \end{equation}

        \begin{equation}
            {\text{SINR}}_d^D = \frac{{{{\left| {{{ { {{{\bf{\hat h}}}_d}{{\bf{B}}_t}} }}{{\bf{w}}_d}} \right|}^2}}}{{\sum\limits_{d' = 1,d' \ne d}^D {{{\left| {{{ {{{{\bf{\hat h}}}_d}{{\bf{B}}_t}} }}{{\bf{w}}_{d'}}} \right|}^2} + {{ { \left({{{\bf{\hat h}}}_d}{{\bf{B}}_t}\right)}^H }}{{\bf{W}}_0}{{{\bf{\hat h}}}_d}{{\bf{B}}_t} + \sigma _d^2} }},\forall d.
        \end{equation}
	\setcounter{equation}{\value{my1}}
\hrulefill
\vspace*{4pt}
\end{figure*}

\subsection{Communication and Radar Sensing SINR
}

 The performance of radar sensing and communication systems is significantly influenced by the respective SINR. In particular, within MIMO radar systems, the detection probability of a point target tends to rise steadily with an increase in the output SINR \cite{6649991}. So we use communication SINR and sensing SINR as the metric for the communication and sensing capabilities of the system. In this paper, we assume that ${\mathbf{v}} \in {\mathbb{C}^{{N_r} \times 1}}$ is the receiving beamforming vector at FD-DFRC-BS to capture the desired reflection sensing signal from the target. For radar sensing target, its sensing SINR can be expressed as Eq. (12), where we define ${\bf{Q}} = \sum\nolimits_{k = 1}^K {{\beta _k}{\bf{A}}\left( {{\theta _k},{\phi _k}} \right)}  + {{\bf{H}}_{{\rm{SI}}}}$. Similarly, we assume $\left\{ {\bf{r}}_u \right\}_{{u = }1}^{U} \in {{\Bbb C}^{{N_r} \times 1}}$ is the receiving beamforming vector at FD-DFRC-BS to obtain uplink transmitted signal. For the uplink, the SINR of the communication data sent by the $u$-th user can be expressed as Eq. (13), where ${\bf{C}} = \sum\nolimits_{k = 0}^K {{\beta _k}{\bf{A}}\left( {{\theta _k},{\phi _k}} \right)}  + {{\bf{H}}_{{\rm{SI}}}} $ represents the interference channel of the downlink path. In addition, for downlink communication, the SINR received by the $d$-th user can be expressed as Eq. (14). For downlink communication, it is assumed that users cannot eliminate the interference from dedicated radar sensing signal ${{\bf{s}}_0}$. In this paper, through the optimization of the beamforming vectors, it can suppress the non-target radar sensing interference and multi-user interference in Eq. (12), the radar snesing interference and multi-user interference in Eq. (13) and Eq. (14).

\subsection{Problem Formulation}
Since MA elements have a certain volume, and different MA elements cannot be in the same candidate discrete position, this paper assumes that the distance from the center of one MA element to the center of another must exceed the specified minimum distance ${D_{\min }}$. Define a matrix ${{\bf{D}}_r} \in {{\Bbb C}^{M \times M}}$, where the element at row $i$ and column $j$ represents the distance between the receiving MA's $i$-th candidate discrete position and its $j$-th candidate discrete position. Hence, the distance between any two receiving MA elements can be articulated as ${\bf{b}}_{r,{n_r}}^T{{\bf{D}}_r}{{\bf{b}}_{r,{n_r}'}}$, ${n_r} \!\ne\! {n_r}'$, $\forall {n_r},{n_r}' \in \left\{ {1,...,{N_r}} \right\}$. According to the definition, let ${{\bf{D}}_t} \in {{\Bbb C}^{N \times N}}$ be defined such that its element in the $i$-th row and $j$-th column represents the distance between the $i$-th transmitting MA element's position and the $j$-th receiving MA element's position. Therefore, the distance between any pair of transmitting MA elements can be expressed as ${\bf{b}}_{t,{n_t}}^T{{\bf{D}}_t}{{\bf{b}}_{t,{n_t}'}}$, ${n_t} \!\ne\! {n_t}'$, $\forall {n_t},{n_t}' \in \left\{ {1,...,{N_t}} \right\}$. For the downlink, the transmit power of FD-DFRC-BS is $\sum\nolimits_{d = 1}^D {||{{\bf{w}}_d}|{|^2}}  + {\rm{Tr}}({{\bf{W}}_0})$. In addition, for the uplink, the transmit power of user is represented as ${p_u}{\text{ = }}{\Bbb E}\left\{ {{{\left| {{d_u}} \right|}^2}} \right\}$, $\forall u$. Therefore, the total transmit power of the system can be expressed as $\sum\nolimits_{d = 1}^D {||{{\bf{w}}_d}|{|^2}}  + {\rm{Tr}}({{\bf{W}}_0}) + \sum\nolimits_{u = 1}^U {{p_u}} $. In this paper, while ensuring the minimum SINR requirements for uplink communication, downlink communication, and radar sensing, we jointly optimize ${{\bf{B}}_r}$, ${{\bf{B}}_t}$, $\left\{ {{{\bf{w}}_d}} \right\}_{d = 1}^D$, ${{\bf{W}}_0}$, $\left\{ {{{\bf{r}}_u}} \right\}_{u = 1}^U$, ${\bf{v}}$ and $\left\{ {{p_u}} \right\}_{u = 1}^U$ to minimize the total transmit power consumption of the system. The specific optimization problem can be formulated as follows
\setcounter{equation}{14}
\begin{subequations}
	\begin{align}
		{\textrm{(P1)}}~~~~&\mathop {\min }\limits_{\scriptstyle{{\bf{B}}_t},{{\bf{B}}_r},\left\{ {{{\bf{w}}_d}} \right\}_{d = 1}^D,{\bf{v}},\hfill\atop
\scriptstyle{{\bf{W}}_0},\left\{ {{p_u}} \right\}_{u = 1}^U,\left\{ {{{\bf{r}}_u}} \right\}_{u = 1}^U\hfill} \sum\limits_{d = 1}^D {||{{\bf{w}}_d}|{|^2} + {\rm{Tr}}\left( {{{\bf{W}}_0}} \right) + \sum\limits_{u = 1}^U {{p_u}} } ,\nonumber \\
		&~~~~~~~~~~{\textrm{s.t.}}\qquad {\text{SIN}}{{\text{R}}^r} \geq {\gamma ^r},\\
		&~~~~~~~~~~~~~~~~~~~{\text{SINR}}_u^U \geq \gamma _u^U,\forall u{,}\\
		&~~~~~~~~~~~~~~~~~~~{\text{SINR}}_d^D \geq \gamma _d^D,\forall d{,}\\
		&~~~~~~~~~~~~~~~~~~~{b_{r,{n_r}}}\left[ m \right] \!\in\! \left\{ {0,1} \right\},\\
		&~~~~~~~~~~~~~~~~{b_{t,{n_t}}}\left[ n \right]\! \in\! \left\{ {0,1} \right\},\\
		&~~~~~~~~~~~~~~~~\sum\limits_{m = 1}^M {{b_{r,{n_r}}}\left[ m \right]}  = 1,\\
		&~~~~~~~~~~~~~~~~\sum\limits_{n = 1}^N {{b_{t,{n_t}}}\left[ n \right]}  = 1,\\
        &~~~~~~~~~~~~~~~~{\bf{b}}_{r,{n_r}}^T{{\bf{D}}_r}{{\bf{b}}_{r,{n_r}'}} \!\geq \!{D_{\min }},{n_r}\! \ne \!{n_r}',\\
        &~~~~~~~~~~~~~~~~{\bf{b}}_{t,{n_t}}^T{{\bf{D}}_t}{{\bf{b}}_{t,{n_t}'}}\! \geq\! {D_{\min }},{n_t}\! \ne \!{n_t}',
	\end{align}
\end{subequations}
where constraints (15a)-(15c) are used to ensure the quality-of-service (QoS) for radar sensing, uplink communication and downlink communication. ${\gamma ^r}$, $\gamma _u^U$ and $\gamma _d^D$ are constant thresholds for the minimum SINR requirement for sensing, the minimum SINR requirement for the uplink transmitting user $u$ and the minimum SINR requirement for the downlink transmitting user $d$. Constraints (15d) and (15e) restrict ${{\bf{B}}_r}$ and ${{\bf{B}}_t}$ to be binary matrices. Constraints (15f) and (15g) ensure that each MA element is only in one candidate discrete position. Constraints (15h) and (15i) ensure that the inter-center separation for any two MA elements is in excess of the minimum allowable distance ${D_{\min }}$. Note that any two different MA elements cannot be in the same candidate discrete position. Due to SINR constraints and the existence of binary variables, it is a mixed-integer non-convex optimization problem, which is hard to obtain the optimal solutions. Next, we will propose a joint optimization framework based on BPSO to solve it.

\section{Joint Discrete Antenna Positioning and Beamforming Optimization Algorithm}
In this section, we propose a joint optimization algorithm to address the problem (P1). We first establish a framework based on the BPSO algorithm in subsection A. Within this framework, we update $\left\{ {{{\bf{w}}_d}} \right\}_{d = 1}^D$, ${{\bf{W}}_0}$ and $\left\{ {{p_u}} \right\}_{u = 1}^U$ according to the discrete candidate positions of MA by solving the fitness function. Since $\left\{ {{{\bf{r}}_u}} \right\}_{u = 1}^U$ and $\bf{v}$ do not directly affect the value of the fitness function, we transform and rewrite $\left\{ {{{\bf{r}}_u}} \right\}_{u = 1}^U$ and $\bf{v}$ with $\left\{ {{{\bf{w}}_d}} \right\}_{d = 1}^D$, ${{\bf{W}}_0}$ and $\left\{ {{p_u}} \right\}_{u = 1}^U$ in subsection B. Since the fitness function is non-convex, we use DC programming and SCA to transform the problem into a convex problem and solve in subsection C. Then, based on the value of fitness function, we search for the current locally and globally optimal solutions for the MA candidate discrete positions to update the MA candidate discrete positions. In turn, we solve for the corresponding $\left\{ {{{\bf{w}}_d}} \right\}_{d = 1}^D$, ${{\bf{W}}_0}$ and $\left\{ {{p_u}} \right\}_{u = 1}^U$ based on the new MA candidate discrete positions. When the iteration is complete, we can obtain the MA candidate discrete positions and their corresponding $\left\{ {{{\bf{w}}_d}} \right\}_{d = 1}^D$, ${{\bf{W}}_0}$ and $\left\{ {{p_u}} \right\}_{u = 1}^U$. At last, the value of $\left\{ {{{\bf{r}}_u}} \right\}_{u = 1}^U$ and $\bf{v}$ can be determined based on the value of $\left\{ {{{\bf{w}}_d}} \right\}_{d = 1}^D$, ${{\bf{W}}_0}$ and $\left\{ {{p_u}} \right\}_{u = 1}^U$.

\subsection{BPSO Algorithm Framework}
Traditional alternating position optimization methods involve fixing the positions of all other MA elements and only moving one to alternate. However, since the candidate positions of MA elements are discrete, there is a possibility of converging to an undesired suboptimal solution. Moreover, as the solution space grows, the computational load increases exponentially. The BPSO algorithm reduces the computational load through the collaborative efforts of discrete particles in a swarm, allowing for a faster approach to the suboptimal solution {\cite{10663427,10464791}}. { Moreover, as the MA related variables are merely coupled within the constraints and exert no direct influence on transmit power, the conventional alternating optimization (AO) framework fails to guarantee a closed feasible region during its iterative alternation.} Taking all factors into account, the BPSO algorithm framework is applied to solve the problem (P1) formulated in this paper. Specifically, taking the receiving MA as an example, BPSO algorithm workflow can be expressed as follows.

\textit{Initialization Position and Speed:} In order to facilitate subsequent calculations, we first assume ${{\bf{\tilde b}}_{r,i}} = {\left[ {{\bf{b}}_{_{r,1}}^T,{\bf{b}}_{_{r,2}}^T, \ldots ,{\bf{b}}_{_{r,{N_r}}}^T} \right]^T} \in {{\Bbb C}^{M{N_r} \times 1}}$. The BPSO algorithm initializes the positions of $I$ particles to ${{\cal B}^{\left( 0 \right)}} = \left\{ {{\bf{\tilde b}}_{r,1}^{\left( 0 \right)},{\bf{\tilde b}}_{r,2}^{\left( 0 \right)}, \ldots ,{\bf{\tilde b}}_{r,I}^{\left( 0 \right)}} \right\}$, where each particle represents a possible distribution of the position of the ${N_r}$ receiving MA elements in $M$ candidate discrete positions. The velocity of $I$ particles is initialized as ${{\cal V}^{\left( 0 \right)}} = \left\{ {{\bf{v}}_1^{\left( 0 \right)},{\bf{v}}_2^{\left( 0 \right)}, \ldots ,{\bf{v}}_I^{\left( 0 \right)}} \right\}$, where ${\bf{v}}_i^{\left( 0 \right)} = {\left[ {{\bf{v}}_{_{r,1}}^T,{\bf{v}}_{_{r,2}}^T, \ldots ,{\bf{v}}_{_{r,{N_r}}}^T} \right]^T} \in {{\Bbb C}^{M{N_r} \times 1}}$ indicates that the speed at which the ${n_r}$-th receiving MA element moves in the $M$ candidate discrete positions, and ${v_{\min }} \leqslant {\bf{v}}_{_{r,{n_r}}}^T\left( m \right) \leqslant {v_{\max }}$.

\textit{Locally Optimal Position and Globally Optimal Position:} Let ${\bf{\tilde b}}_{r,i}^*$ be the locally optimal position of the $i$-th particle. For the $i$-th particle, this position is the optimal solution to the fitness function that it has encountered in its history. Let ${\bf{\tilde b}}_r^ * $ be the {globally} optimal position, which is the optimal solution to the fitness function found among the locally optimal positions of all $I$ particles in the swarm. 

\textit{Velocity Update Criterion:} In the $j$-th iteration, the velocity update of each particle is given by 
\begin{equation}
\begin{aligned}
       {\bf{v}}_i^{\left( j \right)}\left( d \right) &= \omega {\bf{v}}_i^{\left( {j - 1} \right)}\left( d \right) + {c_1}{e_1}\left( {{\bf{\tilde b}}_{r,i}^ * \left( d \right) - {\bf{\tilde b}}_{r,i}^{\left( {j - 1} \right)}\left( d \right)} \right) \\&+ {c_2}{e_2}\left( {{\bf{\tilde b}}_r^ * \left( d \right) - {\bf{\tilde b}}_{r,i}^{\left( {j - 1} \right)}\left( d \right)} \right),
\end{aligned}
\end{equation}
where $j$ is the number of iterations and $0\leq\textit{j}\leq\textit{J}$. $\omega $ is the inertia weight, which can be expressed as 
\begin{equation}
    \omega  = \left( {{\omega _{\max }} - {\omega _{\min }}} \right)\left( {J - j} \right)/J + {\omega _{\min }},
\end{equation}
with ${\omega _{\max }} = 1.2$, ${\omega _{\min }} = 0.4$ generally taken. ${c_1}$ and ${c_2}$ are locally and {globally} learning factors that push each particle towards the locally and {globally} optimal positions, respectively. ${e_1}$ and ${e_2}$ are uniformly distributed random numbers in the range $[0,1]$. They are used to increase randomness and reduce the possibility of converging to an unexpected local optimum.

\textit{Position Update Criterion:} Given the method of probability mapping, we use the sigmoid function to map speed to $[0,1]$ as the probability. This probability is the likelihood that the particle will take a value of 1 in the next step and can be expressed as
\begin{equation}
    s\left( {{\bf{v}}_i^{\left( j \right)}\left( d \right)} \right) = \frac{1}{{1 + {e^{\left( { - {\bf{v}}_i^{\left( j \right)}\left( d \right)} \right)}}}}.
\end{equation}

\textit{Absolute Probability of Position Change:} The position of the receiving MA element is updated according to the probability of the candidate discrete position. Since different MA element positions cannot be at the same candidate discrete position, the position update can be expressed as
\begin{equation}
    \begin{aligned}
        {\bf{\tilde b}}_{r,i}^{\left( j \right)}\!\left( d \right) \!\!=\!\! \left\{ \begin{array}{l}
        1,s\left( {{\bf{v}}_i^{\left( j \right)}\left( d \right)} \right) \ge s\left( {{\bf{v}}_i^{\left( j \right)}\left( k \right)} \right),\\~~\forall k \!\subseteq\! \left[ {\left( {d\! -\! d\% {M}\!+\!1} \right),\!\left( {d\! +\! {M}\! -\! d\% {M}} \right)} \right],\\
        0,{\rm{others}}{\rm{.}}
        \end{array} \right.
    \end{aligned}
\end{equation}

\textit{Fitness Function:} During each iteration, the locally and {globally} optimal positions are updated based on the fitness function value. Assuming the positions meet the constraint conditions, but considering constraints (15h) and (15i). A penalty term is added in the problem (P1), which can be represented as the problem (P2) as follows
\begin{equation}
	\begin{aligned}
		{\textrm{(P2)}}\!\!\!&\mathop {\min }\limits_{\scriptstyle \left\{ {{{\bf{w}}_d}} \right\}_{d = 1}^D,{\bf{v}},{{\bf{W}}_0}, \hfill \atop 
        \scriptstyle \left\{ {{p_u}} \right\}_{u = 1}^U,\left\{ {{{\bf{r}}_u}} \right\}_{u = 1}^U \hfill} \!\! \sum\limits_{d = 1}^D ||{{\bf{w}}_d}|{|^2}\! + \!{\text{Tr}}\left( {{{\bf{W}}_0}} \right) \! +\!\! \sum\limits_{u = 1}^U {{p_u}} \!\! +\! \kappa F\!\left( {{\bf{\tilde t}}_i^{\left( j \right)}} \right)\!,\\
	&~~~~~~~~{\textrm{s.t.}}\qquad 
        \text{(15a)-(15c)},    
	\end{aligned}
\end{equation}
where $F \left( {{\bf{\tilde t}}_i^{\left( j \right)}} \right)$ is a penalty term function that returns the number of MA elements that
violate the minimum MA distance constraint at position ${\bf{\tilde t}}$, ${\bf{\tilde t}}_i^{\left( j \right)} = \left[ {{\bf{t}}_{_{r,1}}^{\text{T}}, \ldots ,{\bf{t}}_{_{r,{N_r}}}^{\text{T}}} \right] \in {{\Bbb C}^{2{N_r} \times 1}},$ and { $\kappa $ is a very large number. Because of the complex SINR constraints, it remains a non-convex optimization problem. When two different MA elements are in the same candidate discrete position, their push particles satisfy constraints (15h) and (15i).} Through iteration, we can generally obtain a suboptimal solution. Generally, as the number of iterations increases, it becomes easier to find a suboptimal solution that is closer to the optimal solution. However, the computational load also increases. How to balance this trade-off is also a potential direction for the future research.

\subsection{Transform of $\left\{ {{{\bf{r}}_u}} \right\}_{u = 1}^U$ and $\bf{v}$ in (P2)}
In this paper, objective function of the problem (P2) is independent of $\left\{ {{{\bf{r}}_u}} \right\}_{u = 1}^U$ and $\bf{v}$. When ${{\bf{B}}_r}$, ${{\bf{B}}_t}$, $\left\{ {{{\bf{w}}_d}} \right\}_{d = 1}^D$, ${{\bf{W}}_0}$ and $\left\{ {{p_u}} \right\}_{u = 1}^U$ are given, $\left\{ {{{\bf{r}}_u}} \right\}_{u = 1}^U$ and $\bf{v}$ only affect ${\text{SINR}}_u^U$ and ${\text{SIN}}{{\text{R}}^r}$ respectively. In order to better realize (P2), the constraints (15a) and (15b) are transformed into
\begin{equation}
    \mathop {{\text{max}}}\limits_{\left\{ {{{\bf{r}}_u}} \right\}_{u = 1}^U} {\text{ SINR}}_u^U,\forall u,
\end{equation}
\begin{equation}
    \mathop {{\text{max}}}\limits_{\bf{v}} {\text{ SIN}}{{\text{R}}^r},
\end{equation}
to solve for $\left\{ {{{\bf{r}}_u}} \right\}_{u = 1}^U$ and $\bf{v}$. In order to facilitate expression, we define 
\begin{equation}
    \begin{aligned}
        ~~~~~~~{{\bf{\Omega }}_u} &= \sum\limits_{u' \ne u}^U {p_{u'}}{{\left( {{{{\bf{\hat g}}}_{u'}}{{\bf{B}}_r}} \right)}^H}{{{\bf{\hat g}}}_{u'}}{{\bf{B}}_r}\\& + {\bf{C}}\left( {\sum\limits_{d = 1}^D {{{\bf{w}}_d}{\bf{w}}_d^H + {{\bf{W}}_0}} } \right){{\bf{C}}^H}  +\sigma _r^2{{\bf{I}}_{{N_r}}} ,\forall u,
    \end{aligned}
\end{equation}
\begin{equation}
    \begin{aligned}
        {\bf{\Theta }}& = \sum\limits_{u = 1}^U {{p_u}{{\left( {{{{\bf{\hat g}}}_u}{{\bf{B}}_r}} \right)}^H}{{{\bf{\hat g}}}_u}{{\bf{B}}_r}} \\& + {\bf{Q}}\left( {\sum\limits_{d = 1}^D {{{\bf{w}}_d}{\bf{w}}_d^H + {{\bf{W}}_0}} } \right){{\bf{Q}}^H} + \sigma _r^2{{\bf{I}}_{{N_r}}}.
    \end{aligned}
\end{equation}
The solutions to optimization problems (21) and (22) are expressed as
\begin{equation}
    {\bf{r}}_u^ *  = {{\bf{\Omega }}_u}^{ - 1}\left({{\bf{\hat g}}_u}{{\bf{B}}_r}\right)^H,\forall u,
\end{equation}
\begin{equation}
    {{\bf{v}}^ * } = {{\bf{\Theta }}^{ - 1}}{{\bf{a}}_r}\left( {{\theta _0},{\phi _0}} \right).
\end{equation}

\subsection{Solution of $\left\{ {{{\bf{w}}_d}} \right\}_{d = 1}^D$,${{\bf{W}}_0}$ and $\left\{ {{p_u}} \right\}_{u = 1}^U$ in (P2)}
By introducing the $\left\{ {{{\bf{r}}_u}} \right\}_{u = 1}^U$, ${\bf{v}}$, ${{\bf{B}}_r}$ and ${{\bf{B}}_t}$ into the corresponding constraints, the problem is transformed into minimizing the total transmit power of the system by optimizing variables $\left\{ {{{\bf{w}}_d}} \right\}_{d = 1}^D$,${{\bf{W}}_0}$ and $\left\{ {{p_u}} \right\}_{u = 1}^U$. We can rewrite the problem (P2) into the problem (P3) as follows
\begin{equation}
	\begin{aligned}
		{\textrm{(P3)}}~~&\mathop {\min }\limits_{\left\{ {{{\bf{w}}_d}} \right\}_{d = 1}^D,\hfill\atop{{\bf{W}}_0},\left\{ {{p_u}} \right\}_{u = 1}^U} \sum\limits_{d = 1}^D {||{{\bf{w}}_d}|{|^2} \!+\! {\text{Tr}}\left( {{{\bf{W}}_0}} \right)\! +\! \sum\limits_{u = 1}^U {{p_u}} }\!+\!\kappa F\left( {{\bf{\tilde t}}_i^{\left( j \right)}} \right),\\
	&~~~~~{\textrm{s.t.}}\qquad 
        \text{(15a)-(15c)}.     
	\end{aligned}
\end{equation}
Due to the non-convex constraints, the problem (P3) is still a non-convex optimization problem. To solve the problem, we introduce a set of auxiliary variables, ${{\bf{W}}_d} = {{\bf{w}}_d}{\bf{w}}_d^H$, ${{\bf{W}}_d}\succeq 0$, ${\text{rank}}\left( {{{\bf{W}}_d}} \right){\text{ = }}1$, $\forall d$, and further assume that ${\bf{\tilde W}} = \sum\nolimits_{d = 0}^D {{{\bf{W}}_d}} $,${\bf{\tilde W}}\succeq 0$, ${\text{rank}}\left( {{\bf{\tilde W}}} \right) = 1$. Next, by substituting ${\bf{\tilde W}}$, we can rewrite Eq. (23) and Eq. (24) as 
\begin{equation}
    {\bf{\Theta }} = \sum\limits_{u = 1}^U {{p_u}{{\left( {{{{\bf{\hat g}}}_u}{{\bf{B}}_r}} \right)}^H}{{{\bf{\hat g}}}_u}{{\bf{B}}_r}}  + {\bf{Q\tilde W}}{{\bf{Q}}^H} + \sigma _r^2{{\bf{I}}_{{N_r}}},
\end{equation}
\begin{equation}
    {{\bf{\Omega }}_u}\! =\! \!\sum\limits_{u' \ne u}^U {{p_{u'}}{{\left( {{{{\bf{\hat g}}}_{u'}}{{\bf{B}}_r}} \right)}^H}{{{\bf{\hat g}}}_{u'}}{{\bf{B}}_r}\! +\! {\bf{C\tilde W}}{{\bf{C}}^H} \!+\! \sigma _r^2{{\bf{I}}_{{N_r}}}} ,\forall u.
\end{equation}
By introducing this variable into the constraints (15a)-(15c), the problem (P3) can be transformed into the problem (P4) as follows
\begin{subequations}
	\begin{align}
		{\textrm{(P4)}}~~~~&\mathop {\min }\limits_{\left\{{{{\bf{W}}_d}} \right\}_{d = 0}^D,\left\{ {{p_u}} \right\}_{u = 1}^U} {\text{Tr}}\left( {{\bf{\tilde W}}} \right) + \sum\limits_{u = 1}^U {{p_u}}+\kappa F\left( {{\bf{\tilde t}}_i^{\left( j \right)}} \right),\nonumber \\
		{\textrm{s.t.}}\qquad &{\left| {{\beta _0}} \right|^2}{\bf{a}}_t^H\left( {{\theta _0},{\phi _0}} \right){\bf{\tilde W}}{{\bf{a}}_t}\left( {{\theta _0},{\phi _0}} \right){\bf{a}}_r^H\left( {{\theta _0},{\phi _0}} \right)\nonumber\\ &{{\bf{\Theta }}^{ - 1}}{{\bf{a}}_r}\left( {{\theta _0},{\phi _0}} \right) \geq {\gamma ^r},\\
		&{p_u}{{{\bf{\hat g}}}_u}{{\bf{B}}_r}{\bf{\Omega }}_u^{ - 1}{\left( {{{{\bf{\hat g}}}_u}{{\bf{B}}_r}} \right)^H} \geq \gamma _u^U,\forall u,\\
		&\left( {1 + \frac{1}{{\gamma _d^D}}} \right){{{\bf{\hat h}}}_d}{{\bf{B}}_t}{{\bf{W}}_d} {\left( { {{{\bf{\hat h}}}_d}{{\bf{B}}_t}} \right)^H} \nonumber\\
        &\geq {{{\bf{\hat h}}}_d}{{\bf{B}}_t}{\bf{\tilde W}}{\left( { {{{\bf{\hat h}}}_d}{{\bf{B}}_t}} \right)^H}  + \sigma _d^2,\forall d \geq 1,\\
        &{{{\bf{W}}_d}} \succeq  0, \forall d,\\
        &{\text{rank}}\left( {{{\bf{W}}_d}} \right) = 1, \forall d,
	\end{align}
\end{subequations}
where ${\bf{\tilde W}} = \sum\nolimits_{d = 0}^D {{{\bf{W}}_d}} $, ${\text{rank}}\left( {{{\bf{W}}_d}} \right){\text{ = }}1$. Due to constraints (30a), (30b) and (30e), the problem is non-convex. Next, we deal with constraints (30a), (30b) and (30e). Since ${{{\bf{W}}_d}} \succeq  0$, ${\gamma ^r} > 0$, ${\bf{a}}_t^H\left( {{\theta _0},{\phi _0}} \right){\bf{\tilde W}}{{\bf{a}}_t}\left( {{\theta _0},{\phi _0}} \right) > 0$, constraint (30a) can be rewritten as
\begin{equation}
        {\bf{a}}_r^H\left( {{\theta _0},{\phi _0}} \right)\!{{\bf{\Theta }}^{ - 1}}{{\bf{a}}_r}\!\!\left( {{\theta _0},{\phi _0}} \right) \!\!\geq\! \!\frac{{{\gamma ^r}}}{{{{\left| {{\beta _0}} \right|}^2}}}{\left(\! {{\bf{a}}_t^H\left( {{\theta _0},{\phi _0}} \right)\!{\bf{\tilde W}}{{\bf{a}}_t}\!\left( {{\theta _0},{\phi _0}} \!\right)} \!\right)^{ - 1}}.
\end{equation}

Considering that $f\left( {\bf{X}} \right) = {{\bf{f}}^H}{{\bf{X}}^{ - 1}}{\bf{f}}$ is a convex function for ${\bf{X}} \succ 0$, the left-hand side of the inequality is a convex function with respect to ${\bf{\Theta }}$. The right-hand side is a convex function with respect to $\left\{ {{{\bf{W}}_d}} \right\}_{d = 0}^D$. Therefore, Eq. (31) is still an non-convex constraint, which can be approximated by a first-order Taylor expansion at the boundary. For the $i$-th iteration of SCA, we consider the following lower bound 
\begin{equation}
    \begin{aligned}
        {\mathbf{a}}_r^H\left( {{\theta _0},{\phi _0}} \right){{\mathbf{\Theta }}^{ - 1}}{{\mathbf{a}}_r}\left( {{\theta _0},{\phi _0}} \right) \geq\\ {\bf{a}}_r^H\left( {{\theta _0},{\phi _0}} \right){\left( {{{\bf{\Theta }}^{\left( {i - 1} \right)}}} \right)^{ - 1}}{{\bf{a}}_r}\left( {{\theta _0},{\phi _0}} \right)\\ - {\bf{a}}_r^H\left( {{\theta _0},{\phi _0}} \right){\left( {{{\bf{\Theta }}^{\left( {i - 1} \right)}}} \right)^{ - 1}}\left( {{\bf{\Theta }} - {{\bf{\Theta }}^{\left( {i - 1} \right)}}} \right)\\{\left( {{{\bf{\Theta }}^{\left( {i - 1} \right)}}} \right)^{ - 1}}{{\bf{a}}_r}\left( {{\theta _0},{\phi _0}} \right) \triangleq f\left( {{\mathbf{\Theta }},{{\mathbf{\Theta }}^{\left( {i - 1} \right)}}} \right),
    \end{aligned}
\end{equation}
where
\begin{equation}
    \begin{aligned}
        {{\mathbf{\Theta }}^{\left( {i - 1} \right)}} &=\sum\limits_{u = 1}^U {p_u^{\left( {i - 1} \right)}{{\left( {{{{\mathbf{\hat g}}}_u}{{\mathbf{B}}_r}} \right)}^H}{{{\mathbf{\hat g}}}_u}{{\mathbf{B}}_r}} \\& + {\mathbf{Q}}{{\mathbf{\tilde W}}^{\left( {i - 1} \right)}}{{\mathbf{Q}}^H} + \sigma _r^2{{\mathbf{I}}_{{N_r}}},
    \end{aligned}
\end{equation}
${{\bf{\tilde W}}^{\left( {i - 1} \right)}} = \sum\nolimits_{d = 0}^D {{\bf{W}}_d^{\left( {i - 1} \right)}} $. $\left\{ {p_u^{\left( {i - 1} \right)}} \right\}_{u = 1}^U$ and $\left\{ {{\bf{W}}_d^{\left( {i - 1} \right)}} \right\}_{d = 0}^D$ are obtained at the $(i-1)$-th iteration. Therefore, a convex subset of the non-convex constraint (30a) is established as 
\begin{equation}
    f\left( {{\bf{\Theta }},{{\bf{\Theta }}^{\left( {i - 1} \right)}}} \right) \!\!\geq\! \frac{{{\gamma ^r}}}{{{{\left| {{\beta _0}} \right|}^2}}}{\left( {{\bf{a}}_t^H\left( {{\theta _0},{\phi _0}} \right){\bf{\tilde W}}{{\bf{a}}_t}\left( {{\theta _0},{\phi _0}} \right)} \right)^{ - 1}}.
\end{equation}
Similarly, we handle the constraint (15b) by first converting constraint (30b) into
\begin{equation}
    {{\bf{\hat g}}_u}{{\bf{B}}_r}{\bf{\Omega }}_u^{^{ - 1}}{\left( {{{{\bf{\hat g}}}_u}{{\bf{B}}_r}} \right)^H} \geq \frac{{\gamma _u^U}}{{{p_u}}},\forall u.
\end{equation}

By processing with the lower bound obtained from the first-order Taylor expansion iteration, for the $i$-th iteration of SCA, we consider the following lower bound
\begin{equation}
    \begin{aligned}
        {{{\bf{\hat g}}}_u}{{\bf{B}}_r}{\bf{\Omega }}_u^{^{ - 1}}{\left( {{{{\bf{\hat g}}}_u}{{\bf{B}}_r}} \right)^H} \geq 
        {{{\bf{\hat g}}}_u}{{\bf{B}}_r}{\left( {{\bf{\Omega }}_u^{\left( {i - 1} \right)}} \right)^{ - 1}}{\left( {{{{\bf{\hat g}}}_u}{{\bf{B}}_r}} \right)^H} 
        - \\{{{\bf{\hat g}}}_u}{{\bf{B}}_r}{\left( {{\bf{\Omega }}_u^{\left( {i - 1} \right)}} \right)^{ - 1}}\!\!\!\left( {{{\bf{\Omega }}_u} - {\bf{\Omega }}_u^{\left( {i - 1} \right)}} \right){\left( {{\bf{\Omega }}_u^{\left( {i - 1} \right)}} \right)^{ - 1}}\!\!{\left( {{{{\bf{\hat g}}}_u}{{\bf{B}}_r}} \right)^H} \\ \triangleq {f_u}\left( {{{\bf{\Omega }}_u},{\bf{\Omega }}_u^{\left( {i - 1} \right)}} \right),\forall u,
    \end{aligned}
\end{equation}
where
\begin{equation}
    \begin{aligned}
        {\bf{\Omega }}_u^{\left( {i - 1} \right)} &= \sum\limits_{u' \ne u}^U p_{_{u'}}^{\left( {i - 1} \right)}{{\left( {{{{\bf{\hat g}}}_{u'}}{{\bf{B}}_r}} \right)}^H}{{{\bf{\hat g}}}_{u'}}{{\bf{B}}_r} \\&+ {\bf{C}}{{{\bf{\tilde W}}}^{\left( {i - 1} \right)}}{{\bf{C}}^H} + \sigma _r^2{{\bf{I}}_{{N_r}}} ,\forall u,
    \end{aligned}
\end{equation}
is obtained at the $(i-1)$-th  iteration. Therefore, a convex subset of the non-convex constraint in constraint (30b) is given by
\begin{equation}
    {f_u}\left( {{{\bf{\Omega }}_u},{\bf{\Omega }}_u^{\left( {i - 1} \right)}} \right) \geq \frac{{\gamma _u^U}}{{{p_u}}},\forall u.
\end{equation}
Based on the convex approximations in Eq. (34) and Eq. (38), in the $i$-th iteration, the problem (P5) is represented as 
\begin{equation}
	\begin{aligned}
		{\textrm{(P5)}}~~~~&\!\!\!\!\!\!
        \mathop{\operatorname{min} } \limits_{\left\{{{{\bf{W}}_d}} \right\}_{d = 0}^D,\left\{ {{p_u}} \right\}_{u = 1}^U} {\text{Tr}}\left( {{\bf{\tilde W}}} \right) + \sum\limits_{u = 1}^U {{p_u}}+\kappa F\left( {{\bf{\tilde t}}_i^{\left( j \right)}} \right),\\
		&~~~~~~~~~{\textrm{s.t.}}\qquad \text{(34), (38), (30c)-(30e)}.
	\end{aligned}
\end{equation}
Since the existence of rank-1 constraint (30e), the problem (P5) is still non-convex. According to the \textbf{Proposition 1}, it can be equivalently written as a DC function constraint.

\noindent \textbf{Proposition 1:} \textit{For a PSD matrix} ${{{\bf{W}}_d}} \in {{\Bbb C}^{{N_t} \times {N_t}}},{\text{Tr}}\left( {{{\bf{W}}_d}} \right) \geqslant1$, $\forall d$, \textit{the rank-1 constraint can be equivalent to the difference between two convex functions, which can be expressed as} 
\begin{equation}
		\text{rank}\left( {{{\bf{W}}_d}} \right) = 1 \Leftrightarrow {\text{Tr}}\left( {{{\bf{W}}_d}} \right) - {\left\| {{{\bf{W}}_d}} \right\|} = 0,
\end{equation}
{\textit{where} ${\text{Tr}}\left( {{{\bf{W}}_d}} \right) = \sum\nolimits_{i = 1}^{N_t} {{\sigma _i}\left( {{{\bf{W}}_d}} \right)} $, ${\sigma _i}\left( {{{\bf{W}}_d}} \right)$} \textit{represents the $i$-th largest singular value of matrix ${{\bf{W}}_d}$, ${\left\| {{{\bf{W}}_d}} \right\|} = {\sigma _1}\left( {{{\bf{W}}_d}} \right).$ }

According to \textbf{Proposition} \textbf{1}, the problem (P6) can then be formulated as
\begin{equation}
	\begin{aligned}
		{\textrm{(P6)}}~~&\mathop {\operatorname{min} }\limits_{\left\{{{{\bf{W}}_d}} \right\}_{d = 0}^D,\hfill \atop 
        \scriptstyle \left\{ {{p_u}} \right\}_{u = 1}^U} {\text{Tr}}\left( {{\bf{\tilde W}}} \right)\! +\! \sum\limits_{d = 0}^D \rho \left( {{\text{Tr}}\left( {{\bf{ W}}_d} \right)\! -\! {{\left\| {{\bf{ W}}_d} \right\|}}} \right)\! \\&~~~~~~~~~~~~~+ \sum\limits_{u = 1}^U {{p_u}}\! + \!\kappa F\left( {{\bf{\tilde t}}_i^{\left( j \right)}} \right),\\
        &~~~~~~~~{\textrm{s.t.}}\qquad \text{(34), (38), (30c), (30d)},
	\end{aligned}
\end{equation}
where $\rho  > 0$ is a penalty factor. 
Since $-{{\left\| {{\bf{W}}_d} \right\|}}$ is concave, the problem (P6) is still non-convex. It can be solved by iterative optimization-minimization techniques. The main idea is to linearize the quartic term $ - \rho {\left\| {{\bf{ W}}_d} \right\|}$ in the objective function and transform into the problem (P7) as follows
\begin{equation}
    \begin{aligned}{\textrm{(P7)}}~~~~&\!\!\!\!\!\!\!\!\!\!\mathop {\operatorname{min} }\limits_{\left\{{{{\bf{W}}_d}} \right\}_{d = 0}^D,\hfill \atop 
        \scriptstyle  \left\{ {{p_u}} \right\}_{u = 1}^U}\!\!\!\!\!\!{\text{Tr}}\left( {{\bf{\tilde W}}} \right) \!+\!\! \sum\limits_{d = 0}^D\rho  \left\langle {{\text{Tr}}\left( {{\bf{ W}}_d} \right),{\bf{I}} \!-\! \partial {{\left\| {{{{\bf{W}}_d}^{i - 1}}} \right\|}}} \right\rangle \\&~~~~~~~~~+ \sum\limits_{u = 1}^U {{p_u}}+\kappa F\left( {{\bf{\tilde t}}_i^{\left( j \right)}} \right),\\
    &~~~~~{\textrm{s.t.}}~~~~\text{(34), (38), (30c), (30d)},
    \end{aligned}
\end{equation}
where ${{\bf{W}}^{i - 1}_d}$ is the optimal solution of the subproblem at $(i-1)$-th iteration. At this point, the problem (P7) is convex and can be solved efficiently using existing solvers such as CVX \cite{2008CVX}. In addition, by solving the following \textbf{Proposition 2}, we can effectively calculate the gradient $\partial {\left\| {{\bf{W}}_d} \right\|}$\cite{9933731}.

\noindent \textbf{Proposition 2:} \textit{For a given PSD matrix ${\bf{W}}_d \in {{\Bbb C}^{{N_t} \times {N_t}}}$, the gradient $\partial {\left\| {{\bf{W}}_d} \right\|}$ can be calculated as ${{\bf{w}}_d}{\bf{w}}^H_d$, where ${{\bf{w}}_d} \in {{\Bbb C}^{{N_t} \times 1}}$ is the principal eigenvector of the matrix ${\bf{W}}_d$.} 

Therefore, when the penalty term is set to zero, the problem (P7) should lead to a rank-1 solution ${{\bf{W}}^ *_d }$, so we can solve it through Cholesky decomposition ${{\bf{\tilde W}}^ * } = {\bf{w}}^{*}_d{\left({\bf{w}}^*_d\right)^H}$.

\begin{algorithm}[htb]
\caption{BPSO Algorithm Framework to Solve (P1)} 
\label{alg:Framwork} 
\begin{algorithmic}[1] 
\REQUIRE  
$\left\{ {{\bf{w}}_d^{(0)}} \right\}_{d = 1}^D$, ${\bf{W}}_0^{(0)}$ and $\left\{ {p_u^{(0)}} \right\}_{u = 1}^U$ and iteration index of BPSO $j$ = 0.
\ENSURE  
${{\bf{B}}_t}$, ${{\bf{B}}_r}$, $\left\{ {{{\bf{w}}_d}} \right\}_{d = 1}^D$, ${\bf{v}}$, ${{\bf{W}}_0}$, $\left\{ {{p_u}} \right\}_{u = 1}^U$ and $\left\{ {{{\bf{r}}_u}} \right\}_{u = 1}^U.$
\STATE \textbf{repeat}
\STATE Update $j$ = $j$ + 1.
\STATE Given $\left\{ {{\bf{w}}_d^{(j - 1)}} \right\}_{d = 1}^D$, ${\bf{W}}_0^{(j - 1)}$ and $\left\{ {p_u^{(j - 1)}} \right\}_{u = 1}^U$, calculate the value of the fitness function to obtain locally optimal ${\bf{\tilde b}}_{i}^ * $ and the current {globally} optimal ${\bf{\tilde b}}^ * $. Then solve Eq. (16)-(19) and store the intermediate solutions ${\bf{B}}_t^{(j)}$ and ${\bf{B}}_r^{(j)}$.
\STATE Given ${\bf{B}}_t^{(j)}$, ${\bf{B}}_r^{(j)}$, solve the problem (P7) and store the intermediate solutions $\left\{ {{\bf{w}}_d^{(j)}} \right\}_{d = 1}^D$, ${\bf{W}}_0^{(j)}$ and $\left\{ {p_u^{(j)}} \right\}_{u = 1}^U$.
\STATE \textbf{Until} \textit{Convergence}.
\STATE According to the {globally} optimal ${\bf{\tilde b}}^ * $, obtain the corresponding solution ${{\bf{B}}_t^{*}}$, ${{\bf{B}}_r^{*}}$, $\left\{ {{{\bf{w}}_d^{*}}} \right\}_{d = 1}^D$, ${{\bf{W}}_0^{*}}$, $\left\{ {{p_u^{*}}} \right\}_{u = 1}^U$.
\STATE Calculate the receiving beamforming $\left\{ {{{\bf{r}}_u^{*}}} \right\}_{u = 1}^U$ and ${{\bf{v}}^{*}}$ according to Eq. (25) and 
Eq. (26), respectively. 
\end{algorithmic}
\end{algorithm}

\subsection{Convergence and Computational Complexity Analysis}
The joint optimization algorithm based on the BPSO framework proposed in this paper can be summarized as \textbf{Algorithm 1}. In this subsection, we analyze its convergence and computational complexity.  Regarding the convergence of the BPSO framework, the {globally} best position is selected based on the value of the fitness function. This position will only be chosen as the {globally} best position when it corresponds to a better value of the fitness function. Therefore, in this paper, the {globally} best position either remains unchanged or moves to a position where the corresponding value of the fitness function decreases. Considering the actual situation, this paper sets a minimum transmit power limit. Meanwhile, \cite{9934931} has proven the convergence of the BPSO algorithm, and its convergence speed is relatively fast. According to \cite{6891348}, a detailed method is proposed to calculate the computational complexity of solving a convex problem through interior point method quantitative analysis. We get the solution to the problem (P3) with complexity order as $\mathcal{O}\left( {\sqrt {{N_t}D + U} \left( {N_t^6{D^3} + N_t^4{D^2}U + {U^3}} \right)} \right)$. The swarm size of BPSO is $I$ and the maximum number of iterations of BPSO is $J$. Since it serves as the process of solving the fitness function of BPSO algorithm, therefore, the overall complexity of the algorithm is  $\mathcal {O}\left( {IJ\sqrt {{N_t}D + U} \left( {N_t^6{D^3} + N_t^4{D^2}U + {U^3}} \right)} \right)$. {Although BPSO introduces additional computational overhead compared with traditional fixed antenna, the spatial DoFs afforded by MA enable a substantial reduction in the required number of antennas, which downsizing directly diminishes the per-iteration computational load.}

\section{Numerical Results}
\newcommand{\tabincell}[2]{\begin{tabular}{@{}#1@{}}#2\end{tabular}}
\begin{table}
	\begin{center}
		\renewcommand{\arraystretch}{}
		\caption{Simulation Parameters}
		\label{T1}
		{\begin{tabular}{|c|c|}
			\hline
		\textbf{Parameters}&\textbf{Value}\\
			\hline
			\tabincell{c}{Number of candidate \\discrete positions}&$M$=$N$=9\\
			\hline
			{Number of MA elements}&${N_t}$=${N_r}$=2\\
			\hline
			\tabincell{c}{Distance between two adjacent\\ candidate discrete positions}&$l$={0.03 m}\\
			\hline
			{Number of uplink users}&$U$=2\\
			\hline
			{Number of downlink users}&$D$=2\\
			\hline
			\tabincell{c}{Number of paths between \\FD-DFRC-BS and users}&$L$=3\\
			\hline
			\tabincell{c}{Noise powers at the \\FD-DFRC-BS and each user} &\tabincell{c}{$\sigma _r^2$=$\sigma _d^2,\forall d$\\{=-80 dB}}\\
			\hline
			{Residual SI channel power}&$\eta _{{n_r},{n_t}}^\text{SI}$=-100 dB\\
			\hline
			\tabincell{c}{Channel fading factor from \\the FD-DFRC-BS to each user}&\tabincell{c}{$\alpha _{u,{l_p}}$ = $\alpha _{d,{l_p}}$\\\=-100 dB}\\
			\hline
			{Gain of target sensing channel}&${\beta _0}${=-50 
dB}\\
			\hline
  			\tabincell{c}{Amplitude of the interference signal}&${\beta _1}$=${\beta _2}${=-90 dB}\\
			\hline
		\end{tabular}}
	\end{center}
\end{table}
In this section, we demonstrate on the effectiveness of the algorithm through numerical simulations. The candidate discrete positions for transmitting MA elements and receiving MA elements are $M$ = $N$ = 9. Assume that both the transmitting and receiving MAs of the FD-DFRC-BS have ${N_t}$ = ${N_r}$ = 2 elements. The frequency is fixed at 5 GHz, which corresponds to a wavelength of $\lambda $ = 0.06 m. Specifically, the distance between two adjacent candidate discrete positions is $l$ = $\lambda $ / 2 = 0.03 m. {Finer steps can be adopted if higher spatial degrees of freedom are desired.} The FD-DFRC-BS serves $U$ = 2 uplink users and $D$ = 2 downlink users. Additionally, the number of paths between the FD-DFRC-BS and both the uplink and downlink users are $L$ = 3. The noise powers at the FD-DFRC-BS and each downlink user are set to $\sigma _r^2=\sigma _d^2,\forall d$ =-80 dB. We assume the residual self-interference (SI) channel power $\eta _{{n_r},{n_t}}^\text{SI}$ = -110 dB. For simplicity, it is assumed that the channel fading factor from the FD-DFRC-BS to each user is $\alpha _{u,{l_p}}$ = $\alpha _{d,{l_p}}$ = -100 dB. The gain of target sensing channel and the amplitude of the interference signal are set to ${\beta _0}$ = -100 dB and ${\beta _1} = {\beta _2} =$ -180 dB. The specific simulation parameters are shown in Table I \cite{10158711}. We primarily consider the impact of sensing SINR constraints, downlink user communication SINR constraints, uplink user transmit power, the number of transmission paths, and the number of users on total transmit power. Meanwhile, we compare MA-enabled ISAC system with the ISAC system equipped with fixed antenna arrays.

\begin{figure}
\centering
\includegraphics[width=2.8in]{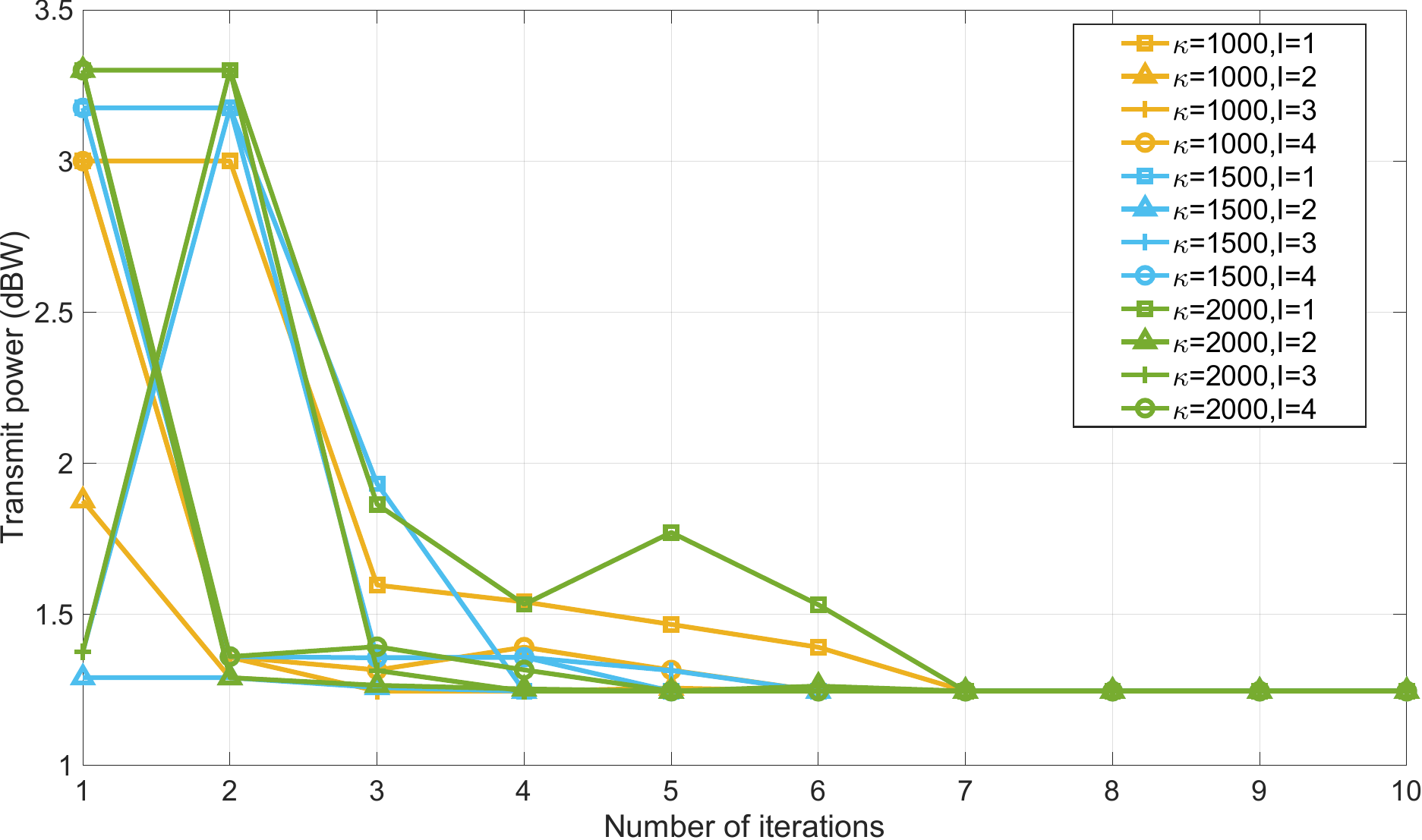}
\caption{Convergence behavior of the proposed algorithm.}
\label{fig2}
\end{figure}

\begin{figure}
\centering
\includegraphics[width=2.8in]{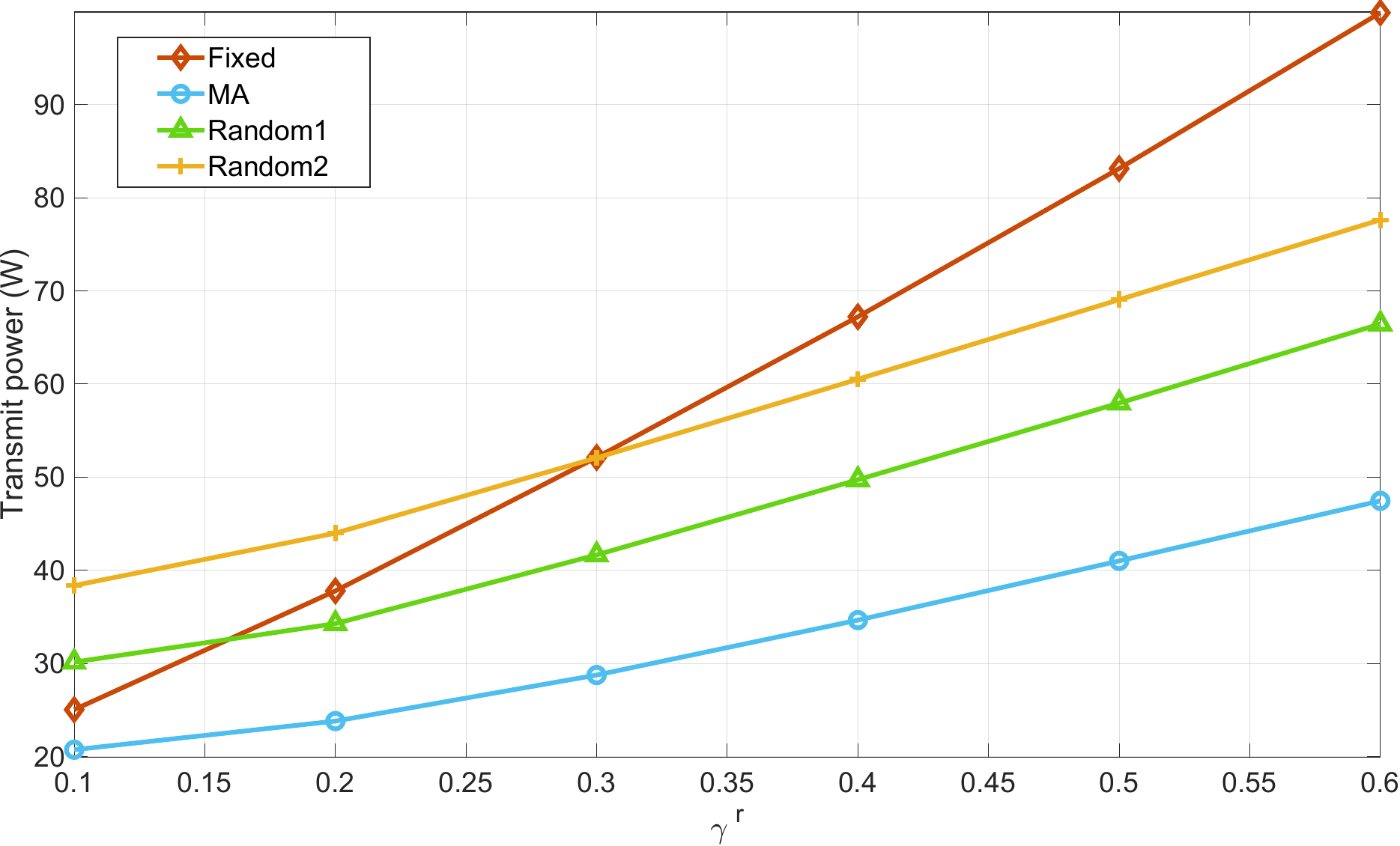}
\caption{Transmit power consumption versus the sensing SINR threshold.}
\label{fig2}
\end{figure}
{Fig. 2 presents the convergence of the proposed algorithm. The penalty factor $\kappa$ prevents any pair of antenna elements from overlapping, while $\rho$ enforces the rank-1 constraint on the beamforming matrix and is therefore set to an extremely large value. Because the objective is to minimize the transmit power, both penalty terms must be sufficiently large to strictly satisfy the physical constraints. If they are too small, the resulting solution will violate the constraints and become meaningless. Experiments are conducted with $\kappa$ = 1000, 1500, and 2000. When $\kappa$ is sufficiently large, the final numerical solutions are identical, and the algorithm converges within six to seven iterations. Convergence is declared when the fitness values of every individual in the population become identical.}

{Fig. 3 describes the relationship between transmit power and the sensing SINR constraint at the FD-DFRC-BS with different types of antennas.} In summary, with a higher minimum necessary sensing SINR, the FD-DFRC-BS expends additional transmit power to fulfill the more rigorous service quality standards expected by the users. In addition, we can observe that within the given range, the fixed linear antenna array, due to suboptimal spatial {DoF}, requires more transmit power compared to the MA. It can be observed that compared with traditional fixed antennas, the MA has a gain of no less than 4.3 W, and the maximum gain of 52.4 W is shown in Fig. 3. Random1 and Random2 are two non-overlapping positions randomly selected from 9 candidate locations, which may lead to poorer channels. Consequently, they need to consume higher transmit power to meet the same sensing requirements. In general, as the sensing SINR increases, the MA-enabled ISAC, due to its superior spatial {DoF}, performs better in minimizing transmit power.

\begin{figure}
\centering
\includegraphics[width=2.8in]{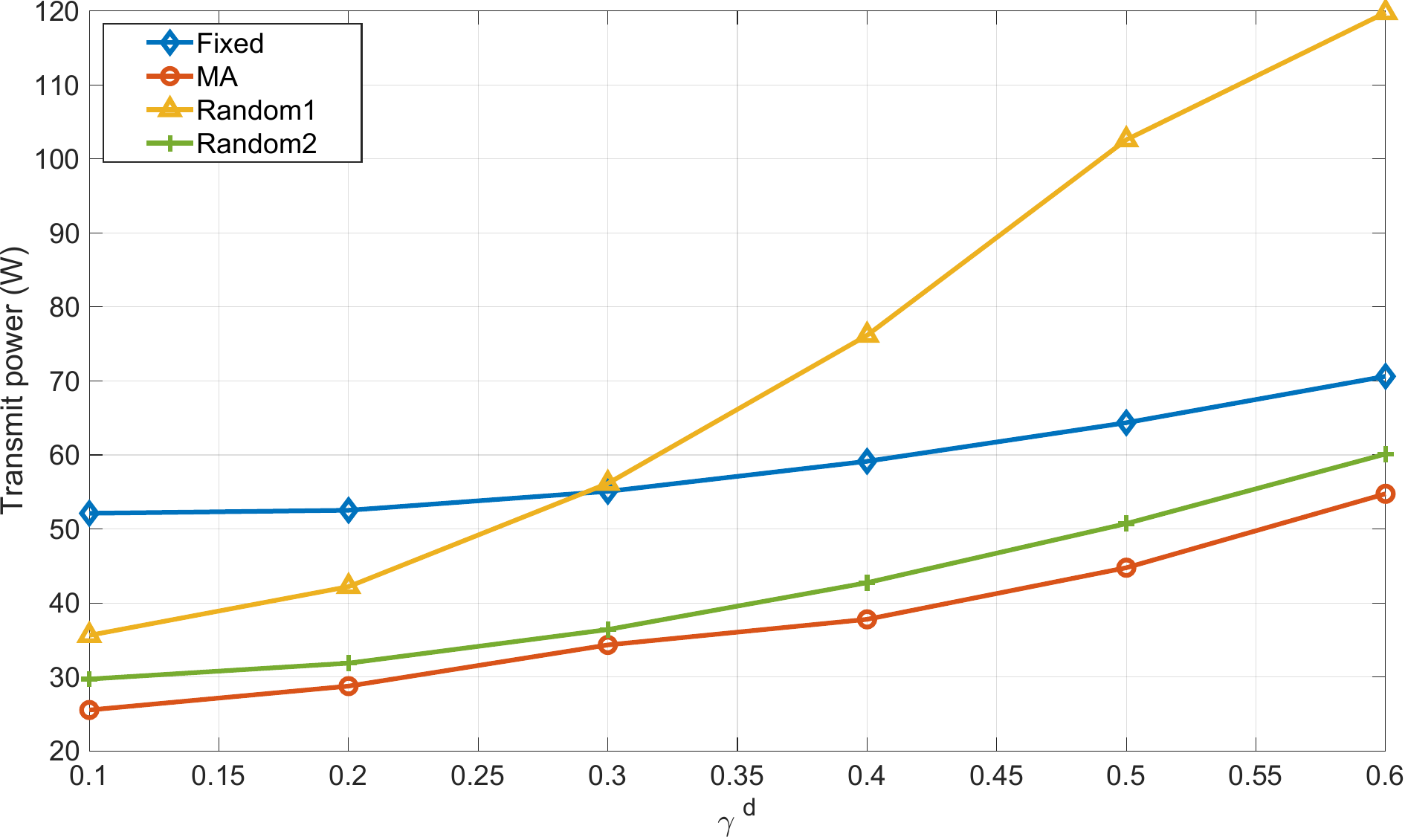}
\caption{Transmit power consumption versus communication SINR threshold.}
\label{fig3}
\end{figure}
{Fig. 4 describes the relationship between transmit power and the downlink user communication SINR constraint at the FD-DFRC-BS with different antenna configurations.} Overall, as the downlink user communication SINR constraint increases, the FD-DFRC-BS consumes more transmit power to meet the stricter quality of service requirements of the users. Additionally, we can observe that within the given range, the fixed linear antenna array, due to suboptimal spatial {DoF}, requires more transmit power compared to the MA, and the transmit power consumption varies significantly with the increase of the downlink user communication SINR constraint. It can be observed that, compared with traditional fixed antennas, MA has a gain of no less than 15.9 W. Random1 and Random2 are two non-overlapping positions randomly selected from 9 candidate locations, which may lead to more random channel quality. Therefore, compared to the MA, they need to consume higher transmit power to meet the same downlink user communication requirements.

\begin{figure}
\centering
\includegraphics[width=2.8in]{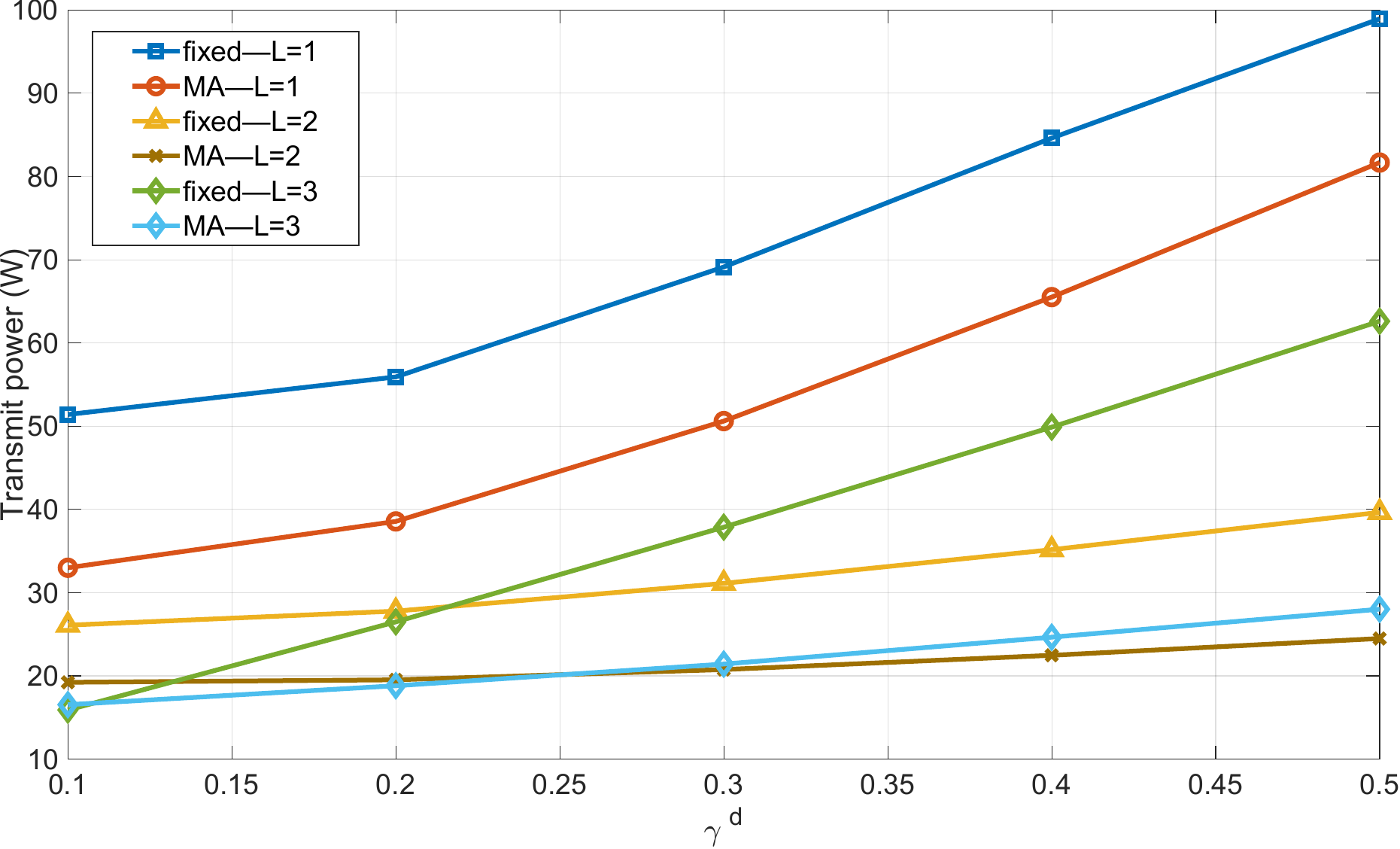}
\caption{Transmit power consumption versus communication SINR threshold for different numbers of transmit paths.}
\label{fig4}
\end{figure}
{Fig. 5 describes the relationship between transmit power and the downlink user communication SINR constraint at the FD-DFRC-BS under different paths.} Overall, as the number of paths decreases, the diversity gain obtained diminishes, hence the FD-DFRC-BS consumes more transmit power to meet the stricter quality of service requirements of the users. Additionally, we can observe that under the same conditions, the fixed linear antenna array, due to suboptimal spatial {DoF}, requires more transmit power compared to the MA with the same number of paths. Specifically, when $L$ = 1, the MA can achieve a maximum gain of 19.1W compared with traditional fixed antennas. When $L$ = 2, the MA can achieve a maximum gain of 15.1W compared with traditional fixed antennas. When $L$ = 3, since the channel is relatively good when $\gamma _d^D,\forall d$ = 0.1, the MA does not show a gain. However, with the change of constraints, the maximum gain can reach 34.6 W. In conclusion, under the same $L$, the gain brought by the MA compared with traditional fixed antennas is relatively significant.

\begin{figure}
\centering
\includegraphics[width=2.8in]{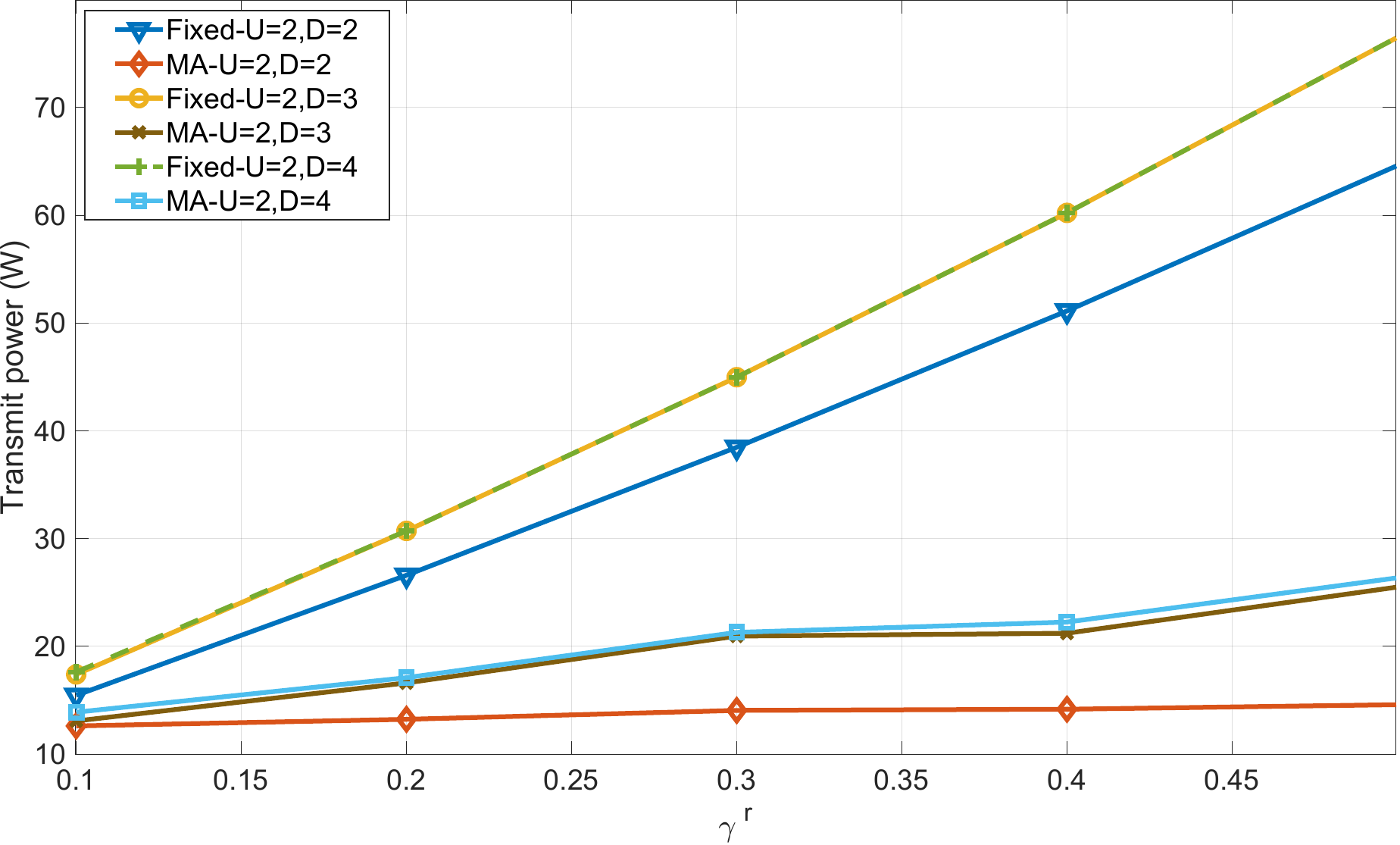}
\caption{Transmit power consumption versus sensing SINR threshold for different numbers of users.}
\label{fig5}
\end{figure}
\begin{figure}
\centering
\includegraphics[width=2.8in]{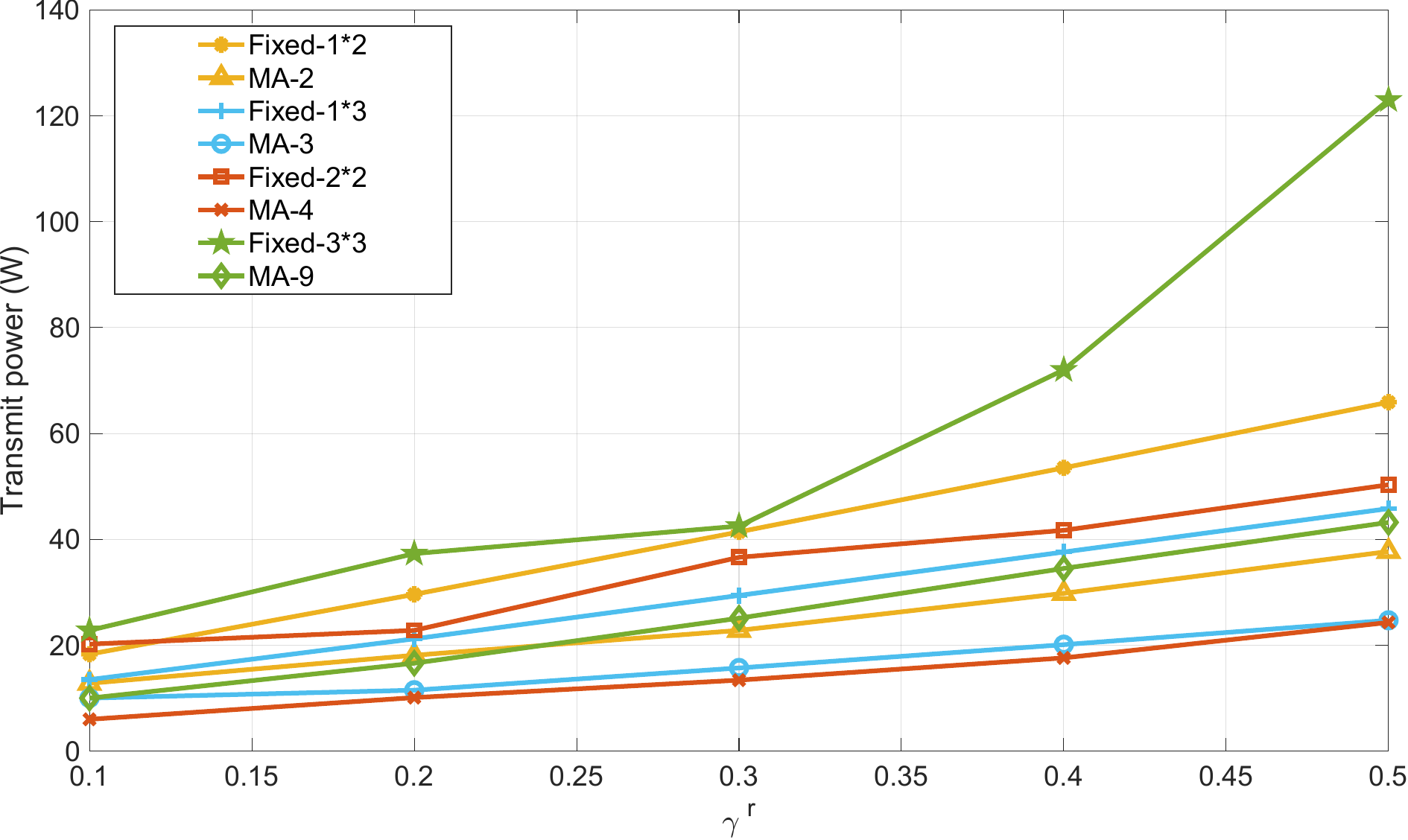}
\caption{Transmit power consumption versus sensing SINR threshold for different numbers of antenna elements.}
\label{fig6}
\end{figure}
\begin{figure}
\centering
\includegraphics[width=2.8in]{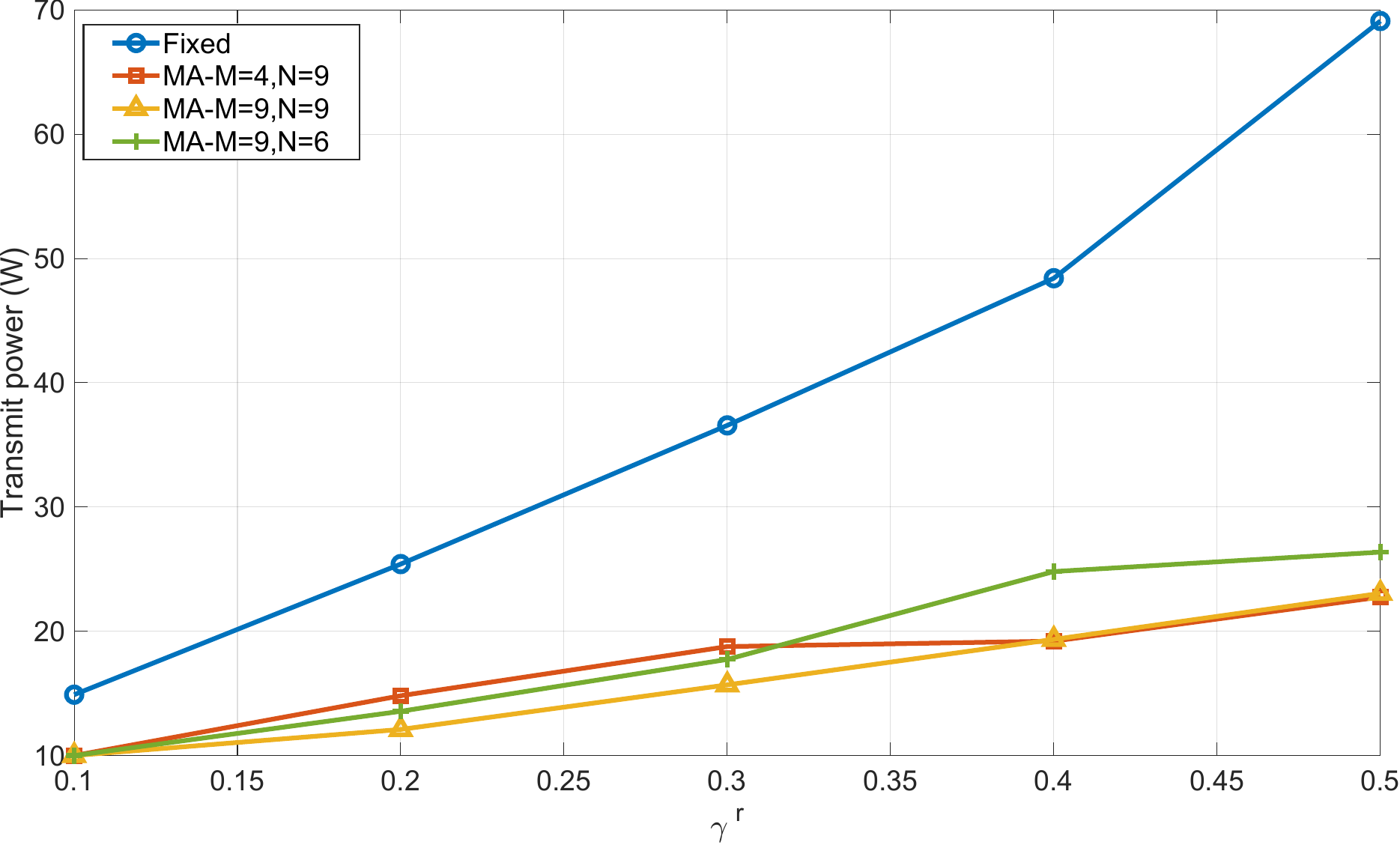}
\caption{Transmit power consumption versus sensing SINR threshold for different numbers of candidate positions.}
\label{fig7}
\end{figure}
{Fig. 6 describes the relationship between the total transmit power and the sensing SINR constraint when the number of users is different.} Overall, it is evident that the transmit power increases progressively as the sensing SINR is raised. Meanwhile, under the same conditions, when the number of downlink users is 2, the transmit power is significantly lower than when there are 3 downlink users. When the number of downlink users changes from 3 to 4, there is no significant increase. By optimizing the beamforming vectors and the positions of MA elements, the system can still allocate appropriate beamforming to each downlink user under limited power. The linear antenna array has higher transmit power in the same scenarios due to its inferior spatial correlation compared to the MA. In contrast, the MA is selected through an algorithm to find the globally suboptimal antenna positions, so under the same circumstances, they consume higher transmit power than the MA. Specifically, when the number of uplink users is consistent, the maximum gain can reach 40.0 W when $D$ = 2, the maximum gain can reach 51.0 W when $D$ = 3, and the maximum gain is 50.2 W when $D$ = 4.

{Fig. 7 describes the relationship between the total transmit power and the sensing SINR constraint when the number of antenna elements is different. When $M$ = $N$ = 16, as the demand for sensing increases, the required minimum transmit power increases. With the same number of antenna elements, MA usually requires less power than fixed antennas due to the spatial freedom brought by the mobility of antenna elements, which shows that MA has an advantage in reducing energy consumption. Specifically, the maximum power reductions achieved by MA are 28.2 W, 17.5 W, 23.2 W, and 79.8 W for configurations with 2, 3, 4, and 9 antenna elements, respectively. In summary, although the minimum transmit power rises with increasing sensing demands across all configurations, MA systems sustain significantly lower power requirements than traditional fixed antennas systems.}

{Fig. 8 describes the relationship the total transmit power and between constraint when the number of candidate discrete positions of the transmitting and receiving antennas is different.} Theoretically, the larger the number of candidate discrete positions of the transmitting and receiving antennas, the more channels are available, and it is more likely to select a more optimal channel, but there is no inevitable relationship between them. As shown in the figure, for MA-M = 9, N = 9 and MA-M = 4, N = 9, when $\gamma _d^r$ = 0.4 and 0.5, the minimum transmit power is almost the same, and their channels are both relatively good. Compared with traditional antennas, they can both obtain a gain of no less than 4.9 W, and the maximum can reach 46.1 W, 45.8 W, and 42.7 W respectively.

\begin{figure}
\centering
\includegraphics[width=3in]{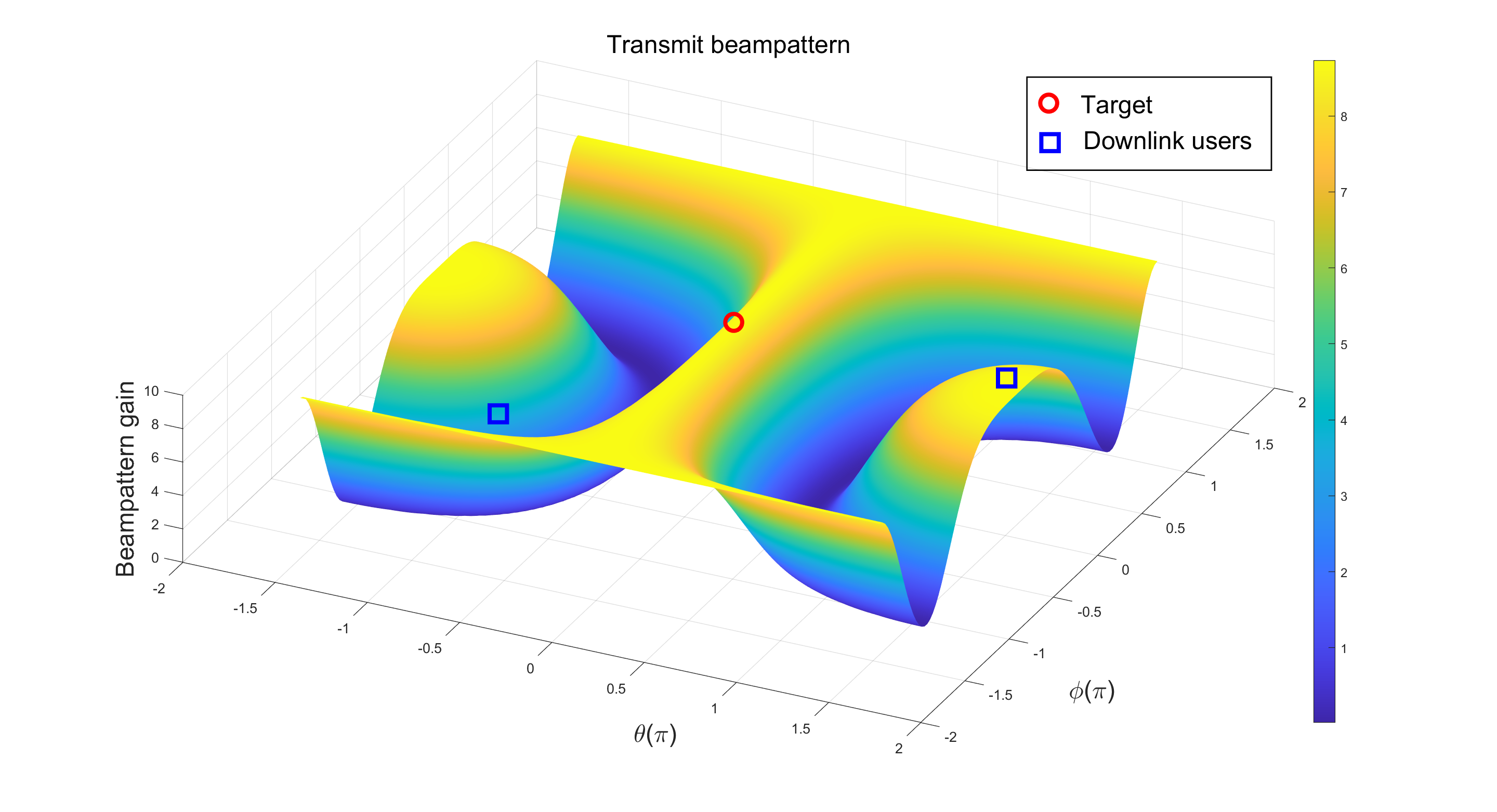}
\caption{Transmit beampattern regarding radar sensing functionality.}
\label{fig8}
\end{figure}
\begin{figure}
\centering
\includegraphics[width=3in]{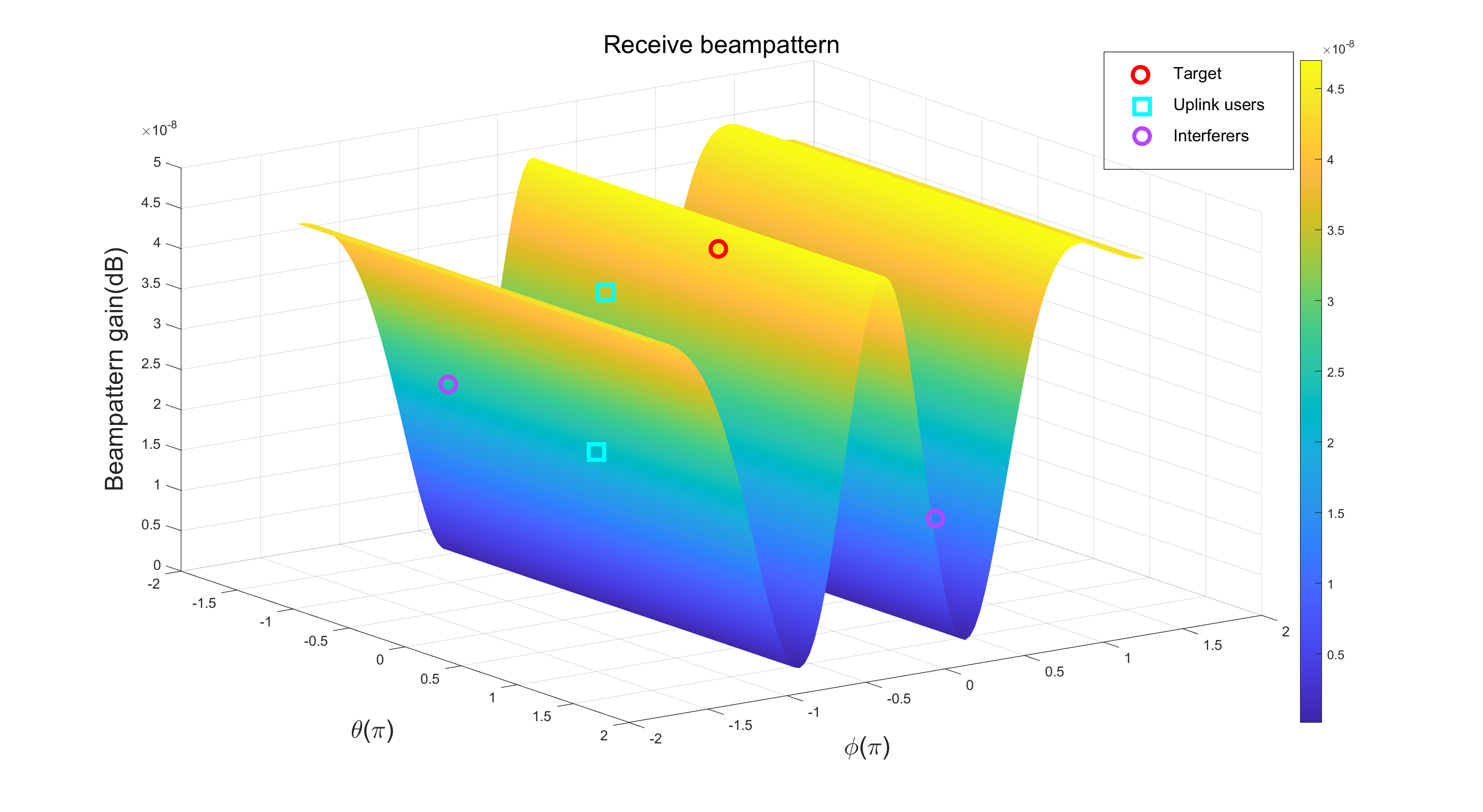}
\caption{Receive beampattern regarding radar sensing functionality.}
\label{fig9}
\end{figure}
Next, we demonstrate the beampattern gain in radar sensing capabilities that is attained through the algorithm. Based on the optimized radar sensing receiving beamforming ${{\bf{v}}^ * }$, which is normalized $\left\| {{{\bf{v}}^ * }} \right\| = 1$ and the transmitted signal ${{\bf{x}}^ * }$, we define the following beampattern
\begin{equation}
    {p_1}\left( {\theta ,\phi } \right){\text{ = }}{\left| {{\bf{a}}_t^H\left( {\theta ,\phi } \right){{\bf{x}}^ * }} \right|^2},
\end{equation}
\begin{equation}
    {p_2}\left( {\theta ,\phi } \right){\text{ = }}{\left| {{{\left( {{{\bf{v}}^ * }} \right)}^H}{{\bf{a}}_r}\left( {\theta ,\phi } \right)} \right|^2}.
\end{equation}
Fig. 9 and Fig. 10 show the two beampatterns achieved by the designed algorithm. Fig. 9 illustrates that the primary transmission beams are oriented towards the target user as well as the downlink users individually. The Fig. 10 shows that when the interference user is relatively close to the uplink user, the interference signal cannot be effectively suppressed due to the insufficiently narrow bandwidth. However, when there is a significant angular separation between the two, the interference signals can be effectively suppressed. Overall, the algorithm is effective for the radar sensing functionality.

\begin{figure}[!t]
\centering
\includegraphics[width=3in]{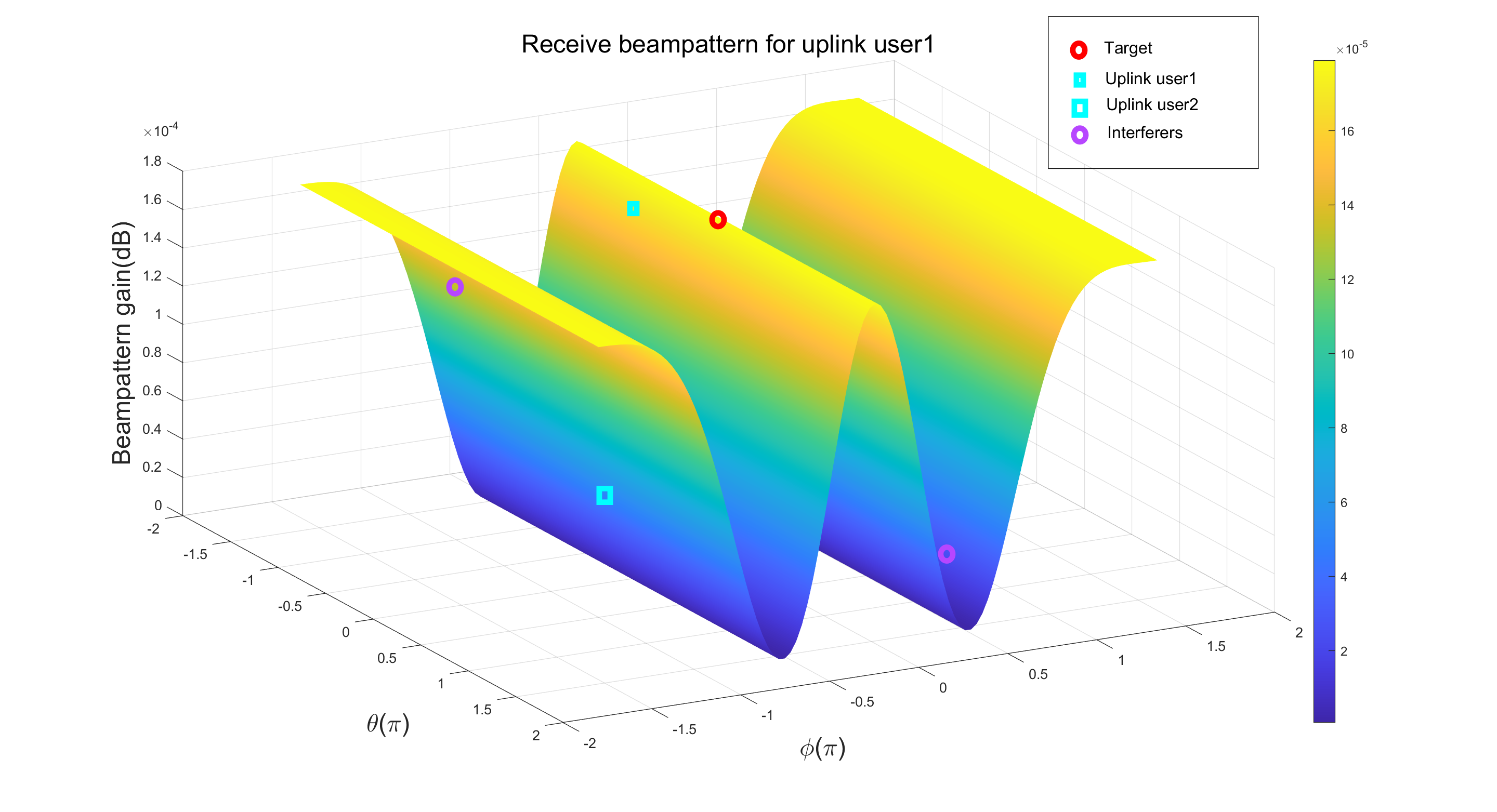}
\caption{Receive beampattern for uplink user 1 regarding communication functionality.}
\label{fig10}
\end{figure}
\begin{figure}[!t]
\centering
\includegraphics[width=3in]{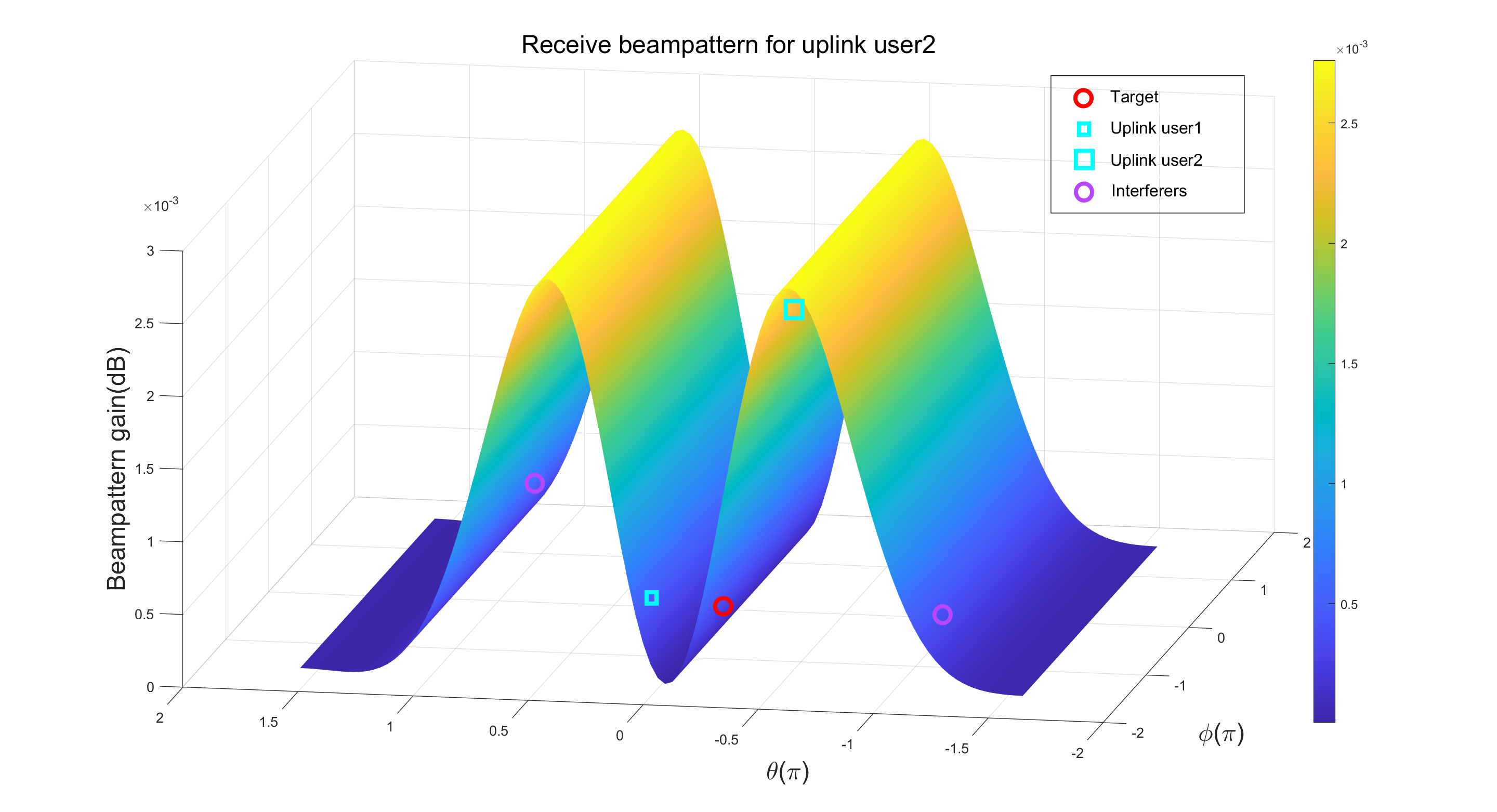}
\caption{Receive beampattern for uplink user 2 regarding communication functionality.}
\label{fig11}
\end{figure}
Fig. 11 and Fig. 12 show the beampattern for communication purposes. Using the optimized communication receiving beamforming, we define the receive beampattern for the uplink user $u$ as ${\left| {{{\left( {{\bf{r}}_u^ * } \right)}^H}{{\bf{a}}_r}\left( {\theta ,\phi } \right)} \right|^2}$. It describes the receive beampattern gain for the two uplink users. From Fig. 11, it can be seen that for the uplink user 1, a main beam pointing towards the user's direction is allocated. Meanwhile, interference user and downlink user are suppressed. Similar observations can also be seen in Fig. 12. Considering the fact that the two main beams of the transmitted signal are directed towards the downlink users, as shown in Fig. 9, it can be concluded that the design is effective in terms of communication functionality.

\section{Conclusions}
This paper has investigated the minimization of transmit power in a full-duplex ISAC system enabled by MA. To solve this problem, we have adopted a framework based on the BPSO algorithm. Initially, the discrete positions of the MA have been determined by iteratively solving the fitness function. For the solution of the fitness function, we have used the DC programming and SCA to handle the non-convex and rank-1 issues within the fitness function. Once the BPSO iteration was completed, the discrete position of the MA elements could be ascertained, and subsequently, the corresponding beamforming vectors, sensing signal covariance matrix, and user transmit power could be obtained or solved. Numerical results have indicated that the system has a performance improvement over traditional ISAC systems. This advantage is mainly due to MA, which increases the spatial {DoF} of the system, allowing the MA-enabled ISAC system to more effectively reduce the total transmit power consumption compared to systems using fixed antenna arrays. In addition, beampattern simulation results also have confirmed that the framework based on the BPSO algorithm was capable of accomplishing a degree of multi-beam alignment and interference suppression.

\balance
\bibliographystyle{IEEEtran}

\bibliography{reference}

@ARTICLE{9540344,
  author={Zhang, J. Andrew and Liu, Fan and Masouros, Christos and Heath, Robert W. and Feng, Zhiyong and Zheng, Le and Petropulu, Athina},
  journal={IEEE J. Sel. Topics Signal Process.}, 
  title={An Overview of Signal Processing Techniques for Joint Communication and Radar Sensing}, 
  year={2021},
  month={Nov.},
  volume={15},
  number={6},
  pages={1295-1315},
  doi={10.1109/JSTSP.2021.3113120}}

@ARTICLE{9585321,
  author={Zhang, J. Andrew and Rahman, Md. Lushanur and Wu, Kai and Huang, Xiaojing and Guo, Y. Jay and Chen, Shanzhi and Yuan, Jinhong},
  journal={IEEE Commun. Surveys \& Tutorials}, 
  title={Enabling Joint Communication and Radar Sensing in Mobile Networks—A Survey}, 
  year={2022},
  month={1st Quar.},
  volume={24},
  number={1},
  pages={306-345},
  doi={10.1109/COMST.2021.3122519}}

@ARTICLE{9737357,
  author={Liu, Fan and Cui, Yuanhao and Masouros, Christos and Xu, Jie and Han, Tony Xiao and Eldar, Yonina C. and Buzzi, Stefano},
  journal={IEEE J. Sel. Areas Commun.}, 
  title={Integrated Sensing and Communications: Toward Dual-Functional Wireless Networks for 6{G} and Beyond}, 
  year={2022},
  month={Jun.},
  volume={40},
  number={6},
  pages={1728-1767},
  doi={10.1109/JSAC.2022.3156632}}

@ARTICLE{7080890,
  author={Björnson, Emil and Matthaiou, Michail and Debbah, Mérouane},
  journal={IEEE Trans. Wireless Commun.}, 
  title={Massive {MIMO} with Non-Ideal Arbitrary Arrays: Hardware Scaling Laws and Circuit-Aware Design}, 
  year={2015},
  month={Aug.},
  volume={14},
  number={8},
  pages={4353-4368},
  doi={10.1109/TWC.2015.2420095}}

@ARTICLE{1266912,
  author={PAULRAJ, A.J. and GORE, D.A. and NABAR, R.U. and BOLCSKEI, H.},
  journal={Proc. IEEE}, 
  title={An overview of {MIMO} communications-a key to gigabit wireless}, 
  year={2004},
  month={Feb.},
  volume={92},
  number={2},
  pages={198-218},
  doi={10.1109/JPROC.2003.821915}}

@ARTICLE{1203154,
  author={Goldsmith, A. and Jafar, S.A. and Jindal, N. and Vishwanath, S.},
  journal={IEEE J. Sel. Areas Commun.}, 
  title={Capacity limits of {MIMO} channels}, 
  year={2003},
  month=jun,
  volume={21},
  number={5},
  pages={684-702},
  doi={10.1109/JSAC.2003.810294}}

@ARTICLE{10938928,
  author={Nguyen, Nhan Thanh and Nguyen, Van-Dinh and Nguyen, Hieu V. and Ngo, Hien Quoc and Swindlehurst, A. Lee and Juntti, Markku},
  journal={IEEE Trans. Signal Process.}, 
  title={Performance Analysis and Power Allocation for Massive {MIMO} {ISAC} Systems}, 
  year={2025},
  month={Mar.},
  volume={73},
  number={},
  pages={1691-1707},
  }

@ARTICLE{2024arXiv240603737S,
    author = {{Singh}, Jitendra and {Srivastava}, Suraj and {Jagannatham}, Aditya K.},
    title = {Energy-efficient hybrid beamforming for integrated sensing and communication enabled mmave {MIMO} systems},
    journal = {arXiv e-prints},
    year = 2024,
    month = jun,
    url = {10.48550/arXiv.2406.03737},
}

@ARTICLE{10557534,
  author={Liao, Bin and Ngo, Hien Quoc and Matthaiou, Michail and Smith, Peter J.},
  journal={IEEE Trans. Wireless Commun.}, 
  title={Power Allocation for Massive {MIMO-ISAC} Systems}, 
  year={2024},
  month={Jun.},
  volume={23},
  number={10},
  pages={14232-14248},
 }

@INPROCEEDINGS{10097000,
  author={Zhu, Minghe and Li, Lei and Xia, Shuqiang and Chang, Tsung-Hui},
  booktitle={Proc. ICASSP 2023-2023 IEEE International Conference on Acoustics, Speech and Signal Processing (ICASSP)}, 
  title={Information and Sensing Beamforming Optimization for Multi-User Multi-Target {MIMO} {ISAC} Systems}, 
  year={2023},
  month={May.},
  address={Rhodes Island, Greece},
  volume={},
  number={},
  pages={1-5},}

@ARTICLE{10770161,
  author={Niu, Yangyang and Wei, Zhiqing and Ma, Dingyou and Yang, Xiaoyu and Wu, Huici and Feng, Zhiyong and Yuan, Jianhua},
  journal={IEEE Trans. Cogn. Commun. Netw.}, 
  title={Interference Management in {MIMO-ISAC} Systems: A Transceiver Design Approach}, 
  year={2025},
  month={Nov.},
  volume={11},
  number={3},
  pages={1762-1775},
}

@ARTICLE{9968163,
  author={Chu, Jinjin and Liu, Rang and Li, Ming and Liu, Yang and Liu, Qian},
  journal={IEEE Trans. Veh. Technol.}, 
  title={Joint Secure Transmit Beamforming Designs for Integrated Sensing and Communication Systems}, 
  year={2023},
  month={Apr.},
  volume={72},
  number={4},
  pages={4778-4791},
  doi={10.1109/TVT.2022.3225952}}

@ARTICLE{9645576,
  author={Wang, Xinyi and Fei, Zesong and Zhang, J. Andrew and Huang, Jingxuan},
  journal={IEEE Commun. Lett.}, 
  title={Sensing-Assisted Secure Uplink Communications With Full-Duplex Base Station}, 
  year={2022},
  month={Feb.},
  volume={26},
  number={2},
  pages={249-253},
  doi={10.1109/LCOMM.2021.3134258}}

@ARTICLE{9144301,
  author={Chowdhury, Mostafa Zaman and Shahjalal, Md. and Ahmed, Shakil and Jang, Yeong Min},
  journal={IEEE Open J. Commun. Society}, 
  title={6{G} Wireless Communication Systems: Applications, Requirements, Technologies, Challenges, and Research Directions}, 
  year={2020},
  month={Jul.},
  volume={1},
  number={},
  pages={957-975},
  doi={10.1109/OJCOMS.2020.3010270}}

@ARTICLE{1353484,
  author={Hejres, J.A.},
  journal={IEEE Trans. Antennas Propag.}, 
  title={Null steering in phased arrays by controlling the positions of selected elements}, 
  year={2004},
  month={Nov.},
  volume={52},
  number={11},
  pages={2891-2895},
  doi={10.1109/TAP.2004.835128}}

@ARTICLE{8060521,
  author={Basbug, Suad},
  journal={IEEE Antennas and Wireless Propag. Lett.}, 
  title={Design and Synthesis of Antenna Array With Movable Elements Along Semicircular Paths}, 
  year={2017},
  month={Oct.},
  volume={16},
  number={},
  pages={3059-3062},
 }

@INPROCEEDINGS{10437926,
  author={Wu, Yifei and Xu, Dongfang and Ng, Derrick Wing Kwan and Gerstacker, Wolfgang and Schober, Robert},
  booktitle={Proc. GLOBECOM 2023 - 2023 IEEE Global Communications Conference}, 
  title={Movable Antenna-Enhanced Multiuser Communication: Jointly Optimal Discrete Antenna Positioning and Beamforming}, 
  year={2023}, 
  address={Kuala Lumpur, Malaysia},
  volume={},
  number={},
  pages={7508-7513},
  doi={10.1109/GLOBECOM54140.2023.10437926}}

@ARTICLE{10684758,
  author={Tang, Jun and Pan, Cunhua and Zhang, Yang and Ren, Hong and Wang, Kezhi},
  journal={IEEE Trans. Commun.}, 
  title={Secure {MIMO} Communication Relying on Movable Antennas}, 
  year={2025},
  month={Apr.},
  volume={73},
  number={4},
  pages={2159-2175},
  }

@ARTICLE{2024arXiv240710393D,
    author = {{Ding}, Jingze and {Zhou}, Zijian and {Jiao}, Bingli},
    title = "{New paradigm for secure full-duplex transmission: Movable antenna-aided multi-user systems}",
    journal = {arXiv e-prints},
    year = 2024,
    month = jul,
    url = {10.48550/arXiv.2407.10393},
}

@ARTICLE{10243545,
  author={Ma, Wenyan and Zhu, Lipeng and Zhang, Rui},
  journal={IEEE Trans. Wireless Commun.}, 
  title={{MIMO} Capacity Characterization for Movable Antenna Systems}, 
  year={2024},
  month={Apr.},
  volume={23},
  number={4},
  pages={3392-3407},
  doi={10.1109/TWC.2023.3307696}}

@ARTICLE{10286328,
  author={Zhu, Lipeng and Ma, Wenyan and Zhang, Rui},
  journal={IEEE Commun. Mag.}, 
  title={Movable Antennas for Wireless Communication: Opportunities and Challenges}, 
  year={2024},
  month={Jun.},
  volume={62},
  number={6},
  pages={114-120},
  doi={10.1109/MCOM.001.2300212}}

@ARTICLE{10278220,
  author={Zhu, Lipeng and Ma, Wenyan and Zhang, Rui},
  journal={IEEE Commun. Lett.}, 
  title={Movable-Antenna Array Enhanced Beamforming: Achieving Full Array Gain With Null Steering}, 
  year={2023},
  month={Dec.},
  volume={27},
  number={12},
  pages={3340-3344},
}

@ARTICLE{10654366,
  author={Liu, Wenchao and Zhang, Xuhui and Xing, Huijun and Ren, Jinke and Shen, Yanyan and Cui, Shuguang},
  journal={IEEE Wireless Commun. Lett.}, 
  title={{UAV}-Enabled Wireless Networks With Movable-Antenna Array: Flexible Beamforming and Trajectory Design}, 
  year={2024},
  volume={},
  number={},
  pages={1-1},
  doi={10.1109/LWC.2024.3451246}}

@ARTICLE{10694747,
  author={Gao, Ying and Wu, Qingqing and Chen, Wen},
  journal={IEEE Trans. Wireless Commun.}, 
  title={Joint Transmitter and Receiver Design for Movable Antenna Enhanced Multicast Communications}, 
  year={2024},
  month={Dec.},
  volume={23},
  number={12},
  pages={18186-18200},
  }

@ARTICLE{2024arXiv240320025D,
    author = {{Ding}, Jingze and {Zhou}, Zijian and {Wang}, Chenbo and {Li}, Wenyao and {Lin}, Lifeng and {Jiao}, Bingli},
    title = "{Secure full-duplex communication via movable antennas}",
    journal = {arXiv e-prints},
    year = 2024,
    month = mar,
    url = {10.48550/arXiv.2403.20025},
}

@ARTICLE{2024arXiv240703228W,
    author = {{Wu}, Haisu and {Ren}, Hong and {Pan}, Cunhua and {Zhang}, Yang},
    title = "{Movable antenna-enabled RIS-aided integrated sensing and communication}",
    journal = {arXiv e-prints},
    year = 2024,
    month = jul,
    url = {10.48550/arXiv.2407.03228},
}

@INPROCEEDINGS{10946353,
  author={Lyu, Wanting and Yang, Songjie and Xiu, Yue and Zhang, Zhongpei and Assi, Chadi and Yuen, Chau},
  booktitle={in Proc. 2024 IEEE 24th International Conference on Communication Technology (ICCT)}, 
  title={Flexible Beamforming for Movable Antenna-Enabled Integrated Sensing and Communication}, 
  year={2024},
  month={Apr.},
  address={Chengdu, China},
  volume={},
  number={},
  pages={1315-1320},
  }

@article{ma2025movableantennaenhancedintegrated,
      title={Movable Antenna Enhanced Integrated Sensing and Communication Via Antenna Position Optimization}, 
      author={Wenyan Ma and Lipeng Zhu and Rui Zhang},
      year={2025},
      month=jan,
      journal={arXiv e-prints},
      eprint={2501.07318},
      archivePrefix={arXiv},
      primaryClass={cs.IT},
      url={https://arxiv.org/abs/2501.07318}, 
}

@ARTICLE{10696953,
  author={Qin, Haoran and Chen, Wen and Wu, Qingqing and Zhang, Ziheng and Li, Zhendong and Cheng, Nan},
  journal={IEEE Wireless Commun. Lett.}, 
  title={Cramér-Rao Bound Minimization for Movable Antenna-Assisted Multiuser Integrated Sensing and Communications}, 
  year={2024},
  month={Dec.},
  volume={},
  number={},
  pages={1-1},}

@INPROCEEDINGS{7360379,
  author={Zhuravlev, Andrey and Razevig, Vladimir and Ivashov, Sergey and Bugaev, Alexander and Chizh, Margarita},
  booktitle={Proc. 2015 IEEE International Conference on Microwaves, Communications, Antennas and Electronic Systems (COMCAS)}, 
  title={Experimental simulation of multi-static radar with a pair of separated movable antennas}, 
  year={2015},
  address={Tel Aviv, Israel},
  volume={},
  number={},
  pages={1-5},
  }

@ARTICLE{6649991,
  author={Cui, Guolong and Li, Hongbin and Rangaswamy, Muralidhar},
  journal={IEEE Trans. Signal Process.}, 
  title={{MIMO} Radar Waveform Design With Constant Modulus and Similarity Constraints}, 
  year={2014},
  month={Jan.},
  volume={62},
  number={2},
  pages={343-353},
  }

@INPROCEEDINGS{10978584,
  author={Cao, Songqi and Zhu, Lipeng and Xiao, Zhenyu and Ning, Boyu},
  booktitle={in Proc. 2025 IEEE Wireless Communications and Networking Conference (WCNC)}, 
  title={Channel Estimation for Movable Antenna Aided Wideband Communication Systems}, 
  year={2025},
  month={May},
  address={Milan, Italy},
  volume={},
  number={},
  pages={1-6},
  }

@article{wang2024movableantennaschannelmeasurement,
      title={Movable Antennas: Channel Measurement, Modeling, and Performance Evaluation}, 
      author={Yiqin Wang and Heyin Shen and Chong Han and Meixia Tao},
      year={2024},
      month=sep,
      journal={arXiv e-prints},
      eprint={2409.03386},
      archivePrefix={arXiv},
      primaryClass={cs.IT},
      url={https://arxiv.org/abs/2409.03386}, 
}

@ARTICLE{10318061,
  author={Zhu, Lipeng and Ma, Wenyan and Zhang, Rui},
  journal={IEEE Trans. Wireless Commun.}, 
  title={Modeling and Performance Analysis for Movable Antenna Enabled Wireless Communications}, 
  year={2024},
  month={Jun.},
  volume={23},
  number={6},
  pages={6234-6250},
}

@ARTICLE{9916163,
  author={Lyu, Zhonghao and Zhu, Guangxu and Xu, Jie},
  journal={IEEE Trans. Wireless Commun.}, 
  title={Joint Maneuver and Beamforming Design for {UAV}-Enabled Integrated Sensing and Communication}, 
  year={2023},
  month={Apr.},
  volume={22},
  number={4},
  pages={2424-2440}}

@ARTICLE{9652071,
  author={Liu, Fan and Liu, Ya-Feng and Li, Ang and Masouros, Christos and Eldar, Yonina C.},
  journal={IEEE Trans. Signal Process.}, 
  title={{Cramér-Rao} Bound Optimization for Joint Radar-Communication Beamforming}, 
  year={2022},
  month={Dec.},
  volume={70},
  number={},
  pages={240-253}}

@ARTICLE{9724174,
  author={Chen, Li and Wang, Zhiqin and Du, Ying and Chen, Yunfei and Yu, F. Richard},
  journal={IEEE J. Sel. Areas Commun.}, 
  title={Generalized Transceiver Beamforming for {DFRC} With {MIMO} Radar and {MU-MIMO} Communication}, 
  year={2022},
  month={Jun.},
  volume={40},
  number={6},
  pages={1795-1808},
  doi={10.1109/JSAC.2022.3155515}}

@ARTICLE{9537599,
  author={Tsinos, Christos G. and Arora, Aakash and Chatzinotas, Symeon and Ottersten, Björn},
  journal={IEEE J. Sel. Topics Signal Process.}, 
  title={Joint Transmit Waveform and Receive Filter Design for Dual-Function Radar-Communication Systems}, 
  year={2021},
  month={Nov.},
  volume={15},
  number={6},
  pages={1378-1392},
  doi={10.1109/JSTSP.2021.3112295}}

@ARTICLE{10663427,
  author={Ding, Jingze and Zhou, Zijian and Li, Wenyao and Wang, Chenbo and Lin, Lifeng and Jiao, Bingli},
  journal={IEEE Commun. Lett.}, 
  title={Movable Antenna-Enabled Co-Frequency Co-Time Full-Duplex Wireless Communication}, 
  year={2024},
  month={Oct.},
  volume={28},
  number={10},
  pages={2412-2416},
}

@INPROCEEDINGS{10464791,
  author={Pi, Xiangyu and Zhu, Lipeng and Xiao, Zhenyu and Zhang, Rui},
  booktitle={Proc. 2023 IEEE Globecom Workshops (GC Wkshps)}, 
  title={Multiuser Communications with Movable-Antenna Base Station Via Antenna Position Optimization}, 
  year={2023},
  month={},
  volume={},
  number={},
  pages={1386-1391},
  address={Kuala Lumpur, Malaysia},}

@article{2008CVX,
  title={{CVX}: MATLAB software for disciplined convex programming},
  author={Grant, M. },
  url={http://cvxr.com/cvx},
  year={2008}}

@ARTICLE{9933731,
  author={Le Thi, Hoai An and Luu, Hoang Phuc Hau and Dinh, Tao Pham},
  journal={IEEE Trans. Neural Netw. Learning Syst.}, 
  title={Online Stochastic {DCA} With Applications to Principal Component Analysis}, 
  year={2024},
  month={Oct.},
  volume={35},
  number={5},
  pages={7035-7047},
  doi={10.1109/TNNLS.2022.3213558}}

@ARTICLE{9934931,
  author={Shekhar, Shashank and Subhash, Athira and Srinivasan, Muralikrishnan and Kalyani, Sheetal},
  journal={IEEE Trans. Commun.}, 
  title={Joint Power-Control and Antenna Selection in User-Centric Cell-Free Systems With Mixed Resolution {ADC}}, 
  year={2022},
  month={Dec.},
  volume={70},
  number={12},
  pages={8400-8415},
}

@ARTICLE{6891348,
  author={Wang, Kun-Yu and So, Anthony Man-Cho and Chang, Tsung-Hui and Ma, Wing-Kin and Chi, Chong-Yung},
  journal={IEEE Trans. Signal Process.}, 
  title={Outage constrained robust transmit optimization for multiuser {MISO} downlinks: {Tractable} approximations by conic optimization}, 
  year={2014},
  month={Nov.},
  volume={62},
  number={21},
  pages={5690-5705},
  doi={10.1109/TSP.2014.2354312}}

@ARTICLE{10158711,
  author={He, Zhenyao and Xu, Wei and Shen, Hong and Ng, Derrick Wing Kwan and Eldar, Yonina C. and You, Xiaohu},
  journal={IEEE J. Sel. Areas Commun.}, 
  title={Full-Duplex Communication for {ISAC}: Joint Beamforming and Power Optimization}, 
  year={2023},
  month={Sep.},
  volume={41},
  number={9},
  pages={2920-2936}}

@ARTICLE{11224420,
  author={Dong, Zhenjun and Zhou, Zhiwen and Xiao, Zhiqiang and Zhang, Chaoyue and Li, Xinrui and Min, Hongqi and Zeng, Yong and Jin, Shi and Zhang, Rui},
  journal={IEEE Trans. Wireless Commun.}, 
  title={Movable Antenna for Wireless Communications: Prototyping and Experimental Results}, 
  year={2025},
  month={early access},
  volume={},
  number={},
  pages={1-1},
  }

@ARTICLE{11007274,
  author={Ning, Boyu and Yang, Songjie and Wu, Yafei and Wang, Peilan and Mei, Weidong and Yuen, Chau and Bjornson, Emil},
  journal={IEEE Wireless Commun.}, 
  title={Movable Antenna-Enhanced Wireless Communications: General Architectures and Implementation Methods}, 
  year={2025},
  month={Oct.},
  volume={32},
  number={5},
  pages={108-116}
}

@ARTICLE{10972180,
  author={Li, Zhendong and Ba, Jianle and Su, Zhou and Huang, Jinyuan and Peng, Haixia and Chen, Wen and Du, Linkang and Luan, Tom H.},
  journal={IEEE Wireless Commun.}, 
  title={Movable Antennas Enabled {ISAC} Systems: Fundamentals, Opportunities, and Future Directions}, 
  year={2025},
  month={early access},
  volume={},
  number={},
  pages={1-8},
}

\end{document}